\documentclass[floatfix,showpacs,preprintnumbers,amsmath,amssymb]{revtex4}

\usepackage{graphicx}
\usepackage{dcolumn}
\usepackage{bm}
\begin{document}
\title{Parity Violating Effects in Elastic Electron Deuteron Scattering}
\author{S. Ahmad$^a$}
\email{physics.sh@gmail.com}
\author{S. K. Singh$^a$}
\author{H. Arenh\"ovel$^b$}
\affiliation{$^a$Department of Physics, Aligarh Muslim University,
  Aligarh-202002, India.\\ 
$^b$Institut f\"ur Kernphysik, Johannes Gutenberg-Universit\"at D-55099,
  Mainz, Germany. } 
\date{\today}
\begin{abstract}
The general expressions for parity violation observables in elastic scattering
of polarized and/or unpolarized electrons from unpolarized deuterons are
given and are numerically evaluated for the kinematics of SAMPLE, PVA4 and G0
experiments. The dominant contribution from the interference of $\gamma$ and Z
exchange as well as the smaller contributions from strangeness ($s{\bar{s}}$)
components of the nucleon, parity odd admixtures in the deuteron wave function,
anapole moments and radiative corrections are included and discussed in the 
context of parity violating electron scattering experiments of present interest. 
\end{abstract}
\maketitle 
\section{Introduction}
Parity violating electron scattering of polarized electrons from nucleons and
nuclei provides an important tool to probe the electroweak structure of
hadrons. The dominant contribution to the parity violating observables in
electron scattering processes comes from the interference of the
electromagnetic 
amplitude given by one photon exchange and the weak neutral current amplitude
given by Z exchange. These processes are therefore considered to be ideal for
studying the structure of weak neutral currents~\cite{Be89,Mc89}. A
knowledge of neutral weak currents is of special importance in understanding
the role of sea quarks in the structure of nucleons~\cite{KaM88}. The isoscalar
sector of the weak neutral current contains information about the strangeness
($s{\bar{s}}$) content of the nucleon which is manifest through the
isoscalar vector and axial vector form factors $G_E^S$, $G_M^S$, and $G_A^S$. A
theoretical and experimental study of these form factors provides thus
information about the role of strange sea quarks in the nucleon. The
initial evidence for the nonzero strangeness ($s{\bar{s}}$) content of the
nucleon as seen in deep inelastic scattering (DIS)~\cite{As88}, neutrino
scattering~\cite{Ah87} and pion nucleon scattering~\cite{DoN86} 
has generated great interest in looking for these effects through parity
violating observables in elastic and inelastic scattering of electrons from
nucleons and nuclei. 

Many experimental programs have been started at various
electron accelerators to search for the strangeness content of
the nucleon through the observation of parity violating asymmetries in the
scattering of polarized electrons from proton~\cite{BePS05}-\cite{Ar05},
deuteron~\cite{Sp00},~\cite{It04,Ani06} and
$^4{\text{He}}$~\cite{Ac07,Ani06} targets from which first results
have been reported for the strange form
factors~\cite{LiMR07}-\cite{Pa04}. The initial results for the
electron 
asymmetry in quasi-elastic scattering of polarized electrons from
deuterium targets in the SAMPLE II experiment indicated that the effect of
radiative corrections in the weak axial coupling constant $G_A^S$ could be
large, making it  difficult to extract information on the strangeness charge
and magnetic moment form factors $G_E^S$ and $G_M^S$. Later, the updated results
from SAMPLE II and the new results from SAMPLE III~\cite{It04,Ani06}
have shown that 
the contribution to the asymmetry from radiative
corrections~\cite{Pa04}-\cite{MH90} to the weak coupling constants, especially the
contributions like anapole moments~\cite{ZhPHR00}-\cite{MH90} of the
deuteron etc., are small. 

The effect of P odd admixtures in the wave functions of deuteron
and the two nucleon continuum to the asymmetry is
now shown to be quite small in the kinematic region of these quasi-elastic scattering
experiments~\cite{ScCP03,LiPR03}. However, it was realised that
corrections due to the related processes of elastic electron deuteron
scattering and coherent pion production in the kinematic region of this
experiment could be important. It was also emphasized that quasi-elastic
electron deuteron scattering at backward angles is predominantly sensitive to
$G_A^S$ and thus can be used to determine $G_A^S$. With a better knowledge of
$G_A^S$ from deuterium experiments, the extraction of $G_E^S$ and $G_M^S$ from
hydrogen measurements will be facilitated. In view of this, the backward angle
measurements of the parity violating asymmetry are being done in various
scattering experiments with polarized electrons on  deuterium
targets~\cite{Ba07}-\cite{Ro06}. 

In theory, many calculations have been done for the parity violating
helicity dependent electron asymmetry in quasi-elastic scattering of polarized
electrons off deuterons~\cite{ScCP03,LiPR03} and~\cite{HwHM81}-\cite{PO90}. But
there exist very few calculations for the case of elastic electron deuteron
scattering~\cite{PO90}-\cite{HwH80}. Recently, we have presented a general 
formalism for calculating all parity violating observables in elastic 
scattering of polarized and unpolarized electrons from polarized and/or
unpolarized deuteron targets due to the interference of the weak and
electromagnetic amplitudes~\cite{ArS01}. The contribution of the strangeness
component of 
the nucleon to the electron asymmetry in the scattering of polarized electrons
from deuteron along with other contributions from the P odd parity admixture
in the deuteron due to a parity violating nucleon-nucleon potential, anapole
moments, exchange currents, and radiative corrections to the weak coupling
constants are important theoretical ingredients, which should be studied in
order to analyse the present experiments being done to observe the parity
violating asymmetries in scattering experiments with polarized and unpolarized
electrons. 

In this paper we have studied the parity violating observables in elastic
scattering of polarized and unpolarized electrons from unpolarized deuteron
targets at electron energies relevant for SAMPLE~\cite{It04,Ani06}, PVA4~\cite{Ba07}
and G0~\cite{Ro06} experiments. In
particular, we have calculated the novanishing parity violating parts of
the deuteron vector recoil polarisation for elastic scattering of
unpolarized electrons from unpolarized deuterons. In the case of elastic
scattering of polarized electrons from unpolarized deuterons, we have
calculated the parity violating helicity dependent electron
asymmetry as well as the parity violating nonvanishing components of the
recoil deuteron polarisation. The separate contributions due to the 
$\gamma$-$Z$ interference and the odd parity admixture in the deuteron arising from
from the parity violating nucleon-nucleon potential have been calculated.
The additional contribution due to the
strangeness components of the nucleon, including anapole moments and radiative
corrections to the weak coupling constants have been evaluated. The effect of
meson exchange currents in the vector current contribution has been included 
through the use of Siegert's theorem for calculating the parity
violating electric amplitude  
due to the vector current which is non-vanishing because of the odd parity 
admixture in the deuteron wave function.  

In section II, we describe briefly the basic formalism for calculating all
the parity violating observables in elastic scattering of polarized electrons
from unpolarized  deuterons. In section III and IV, we describe some technical details
regarding the odd parity wave function components of the deuteron and
the electroweak currents needed for calculating various contributions
to parity violating observables. In section V, we present the
numerical results, and a summary of our work with conclusions given in section VI. 

\section{Formalism}
Here we will give a brief review of the salient features of the formal 
expressions for cross section and polarization observables starting 
with a short outline of the classification of the 
various hadronic currents as defined in~\cite{ArS01} whose multipoles enter 
in the expressions for the observables given below in 
subsection~\ref{observable}. 

\subsection{Classification of Currents}\label{currents}
The starting point of the classification 
is the general expression for the invariant scattering matrix elements
containing both contributions from virtual $\gamma$ and $Z$ exchange
\begin{eqnarray}
{\cal
  M}_{fi}=\frac{e^2}{q_\mu^2}\Big(j^{\gamma,\,\mu}_{fi}J^{\gamma}_{fi,\,\mu} 
+\frac{1}{\sin^2(2\theta_W)}\frac{q_\mu^2}{M_Z^2-q_\mu^2}\,
j^{Z,\,\mu}_{fi}J^{Z}_{fi,\,\mu}\Big)\,,
\end{eqnarray}
denoting the electromagnetic and neutral currents of lepton and hadron by
$j^{\gamma/Z}_\mu$ and $J^{\gamma/Z}_\mu$, respectively, and with 
$\theta_W$ the Weinberg angle. The lepton electromagnetic and weak neutral
currents are separated into vector ($j^v$) and axial vector ($jâ$)
contributions according to
\begin{eqnarray}
j^{\gamma,\,\mu}&=&\bar u(k_2)\gamma^\mu\, u(k_1)=j^{v,\,\mu}\,,\\
j^{Z,\,\mu}&=&\bar u(k_2)(g_v\gamma^\mu+g_a\gamma^\mu\gamma_5)\, u(k_1)
=j^{v,\,\mu}+j^{a,\,\mu}\,.
\end{eqnarray}
With respect to the hadron current, it is useful for the evaluation 
to distinguish between the 
contribution which couples to the lepton vector current and the one coupling
to the lepton axial vector current by writing the matrix element in the form
\begin{eqnarray}
{\cal M}_{fi}=\frac{e^2}{q_\mu^2}\Big(j^{v,\,\mu}J_{fi,\,\mu}({\cal V})
+j^{a,\,\mu}J_{fi,\,\mu}({\cal A})\Big)
\end{eqnarray}
where we have introduced
\begin{eqnarray}
J_{fi,\,\mu}({\cal V})&=&J_{fi,\,\mu}^{\gamma}+J_{fi,\,\mu}^{Z^{\cal V}}\,,\\
J_{fi,\,\mu}({\cal A})&=&J_{fi,\,\mu}^{Z^{\cal A}}\,,
\end{eqnarray}
with
\begin{eqnarray}
J_{fi,\,\mu}^{Z^{\cal V/A}}=\widetilde G_{v/a}\,J^Z_{fi,\,\mu}\,,
\end{eqnarray}
where
\begin{eqnarray}
\widetilde G_{v}&=&\frac{g_{v}}{\sin^2(2\theta_W)}\frac{q_\mu^2}{M_Z^2-q_\mu^2}
=\frac{4\sin^2\theta_W-1}{2\sin^2(2\theta_W)}\frac{q_\mu^2}{M_Z^2-q_\mu^2}\,,\\
\widetilde G_{a}&=&\frac{g_{a}}{\sin^2(2\theta_W)}\frac{q_\mu^2}{M_Z^2-q_\mu^2}
=\frac{1}{2\sin^2(2\theta_W)}\frac{q_\mu^2}{M_Z^2-q_\mu^2}\,.
\end{eqnarray}
Here, the arguments ${\cal V}$ and ${\cal A}$ merely indicate to which type of 
lepton current the hadronic current couples. Both current types contain vector
as well as axial contributions. We have further classified in~\cite{ArS01} both
contributions by their vector and axial parts. The electromagnetic current 
consists of only a vector piece while the neutral current contains both 
vector and axial ones, i.e.\ in an obvious notation
\begin{eqnarray}
J^Z_{fi,\,\mu}=J^{Z_v}_{fi,\,\mu}+J^{Z_a}_{fi,\,\mu}\,.
\end{eqnarray}
Thus we have
\begin{eqnarray}
J_{fi,\,\mu}({\cal V/A})&=&J_{fi,\,\mu}^{\gamma}
+\widetilde G_{v/a}(J_{fi,\,\mu}^{Z_{v}}+J_{fi,\,\mu}^{Z_{a}})\nonumber\\
&=&J_{fi,\,\mu}^{\gamma}+J_{fi,\,\mu}^{Z^{\cal V/A}_v}+
J_{fi,\,\mu}^{Z^{\cal V/A}_a}\,.
\end{eqnarray}
Altogether, we have five types of hadronic currents, namely three of vector
type $J_{fi,\,\mu}^{\gamma}$, $J_{fi,\,\mu}^{Z^{\cal V}_v}$, and 
$J_{fi,\,\mu}^{Z^{\cal A}_v}$, and two of axial vector type 
$J_{fi,\,\mu}^{Z^{\cal V}_a}$ and $J_{fi,\,\mu}^{Z^{\cal A}_a}$.

\subsection{Definition of Multipoles}\label{multipoles}

The multipole moments appearing in the expressions of the various observables
below 
are defined by the multipole expansion of the $t$-matrix contributions to the
elastic scattering process arising from the hadronic vector and axial vector 
current contributions according to the classification in the previous
subsection for which we will use a common label ``$c$''
\begin{eqnarray}
t_{m'\lambda m}^c&=&\frac{\sqrt{E_d'E_d}}{M_d}\langle m'|J_\lambda(c)|m\rangle
\nonumber\\
&=&\frac{\sqrt{E_d'E_d}}{M_d}(-)^{\lambda} a_\lambda\sum_L i^L\hat L
\langle 1m'|{\cal O}_{L\lambda}(c)|1m\rangle\nonumber\\
&=&(-)^{1-m'+\lambda} a_\lambda\sum_L i^L\hat L 
\left(\begin{matrix}
1 & L & 1\cr
-m' & \lambda & m\cr
\end{matrix}\right){O}^\lambda_L(c)\,,
\end{eqnarray}
where $a_\lambda=\sqrt{2\pi(1+\delta_{\lambda 0})}$ and the general 
multipole operator is defined by
\begin{eqnarray}
{\cal O}^\lambda_{LM}(c)&=&\delta_{\lambda 0}\,{\cal C}_{LM}(c)
+\delta_{|\lambda|1}({\cal E}_{LM}(c)+\lambda {\cal M}_{LM}(c))\,,
\end{eqnarray}
with ${\cal C}_{LM}$, ${\cal E}_{LM}$, and ${\cal M}_{LM}$ denoting
charge, electric and magnetic multipoles, respectively. 
The corresponding reduced matrix elements between deuteron states are given by
\begin{eqnarray}
{O}^\lambda_L(c)&=&\frac{\sqrt{E_d'E_d}}{M_d}
\langle 1\|{\cal O}^\lambda_L(c)\|1\rangle\nonumber\\
&=&\delta_{\lambda 0}\,{C}_L(c)
+\delta_{|\lambda|1}({E}_{L}(c)+\lambda {M}_{L}(c))\,.
\end{eqnarray}

\subsection{Observables}\label{observable}

We will start from the general expression for a polarization observable 
of elastic electron deuteron scattering
\begin{eqnarray}
e(k_1)+d(d)\rightarrow e(k_2)+d(d')
\end{eqnarray}
including longitudinal electron polarization of degree $h$ but for an
unpolarized deuteron target as derived in~\cite{ArS01} and given there in
Eq.~(123) 
\begin{eqnarray}
{\cal O}_X\frac{d\sigma^{\gamma+Z}}{d\Omega_{k_2}^{\mathrm{lab}}}&=&
\,\sigma_{\mathrm{Mott}}\,S_0\Big[A^0_d(X)+hA^0_{ed}(X)\Big]\,,\label{polobs}
\end{eqnarray}
with
\begin{eqnarray}
\,\sigma_{\mathrm{Mott}}\,=
\frac{\alpha^2\cos^2\frac{\theta_e^{\mathrm{lab}}}{2}}{4\sin^4
\frac{\theta_e^{\mathrm{lab}}}{2}}
\frac{k_2^{\mathrm{lab}}}{(k_1^{\mathrm{lab}})^3}\label{mott}
\end{eqnarray}
for the Mott cross section.

The four-momenta of incoming and scattered electrons are denoted by $k_1$ and 
$k_2$, respectively, and the corresponding deuteron four-momenta by 
$d=(E_d,\vec d\,)$ and $d'=(E_d',\vec d'\,)$, respectively. Furthermore, 
$q_\mu^2=q_0^2-\vec q^{\,2}$ denotes the squared four-momentum transfer with 
$q=k_1-k_2$. The coordinate system chosen is such that the $z$-axis is 
taken along the momentum transfer $\vec q$, the $y$-axis along 
$\vec k_1\times \vec k_2$, i.e.\ pependicular to the scattering plane, 
and the $x$-axis as to form a right-handed system. 

For an observable ${\cal O}_X$ we had introduced in~\cite{ArS01} the 
short-hand notation $X=(IM\pm)$ with $I=0,1,2$ and $I\geq M\geq 0$. In 
detail, for $I=0$ one has only $(00+)$ which refers to the differential 
cross section, $(1M\pm)$ to the deuteron vector and $(2M\pm)$ the tensor 
recoil polarization components in the spherical basis. The cartesian 
components are given for the vector polarization by
\begin{eqnarray}
P_{x/y}= \pm\frac{1}{\sqrt{3}}{\cal O}_{11\pm}\,,\qquad 
P_{z}= \sqrt{\frac{2}{3}}{\cal O}_{10+}\,,
\end{eqnarray}
and for tensor polarization
\begin{eqnarray}
\begin{array}{ll}
P_{xx/yy}= \pm\frac{1}{2\sqrt{3}}{\cal O}_{22+}-\frac{1}{2}P_{zz}\,,\quad
& P_{zz}= \frac{\sqrt{2}}{3}{\cal O}_{20+}\,,\\
 & \\
P_{xy}= -\frac{1}{2\sqrt{3}}{\cal O}_{22-}\,,
& P_{zx/zy}= \mp\frac{1}{2\sqrt{3}}{\cal O}_{21\pm}\,,\\
\end{array}
\end{eqnarray}
where the cartesian components are defined by the deuteron density matrix 
in the form
\begin{eqnarray}
\rho^d=\frac{1}{3}\Big(1+\vec P\cdot \vec S+\sum_{kl}P_{kl}S^{[2]}_{kl}\Big)\,.
\end{eqnarray}
Here $\vec S$ denotes the deuteron spin operator and $S^{[2]}$ the
corresponding irreducible tensor of rank 2 
\begin{eqnarray}
S^{[2]}_{kl}=\frac{1}{2}(S_kS_l+S_lS_k)-\frac{1}{3}\delta_{kl}\,.
\end{eqnarray}
We now will list the various asymmetries of (\ref{polobs}).
\begin{description}
\item{(i)}
Differential cross section:
\begin{eqnarray}
\frac{d\sigma^{\gamma+Z}}{d\Omega_{k_2}^{\mathrm{lab}}}&=&
\,\sigma_{\mathrm{Mott}}\,S_0
\Big[1 + A^{00+}_d(00+)_{pv}+hA^{00+}_{ed}(00+)_{pv}\Big]\,,
\label{diffcross}
\end{eqnarray}
denoting by the subscript ``$pv$'' a parity violating contribution. 
The unpolarized parity conserving differential cross section is given by
$\,\sigma_{\mathrm{Mott}}\,S_0$ with
\begin{eqnarray}
S_0=\frac{4\pi}{3}\Big(((C_0^\gamma)^2+ (C_2^\gamma)^2)v_L 
+(M_1^\gamma)^2v_T\Big)\,,
\end{eqnarray}
and the parity violating asymmetries by
\begin{eqnarray}
S_0A^{00+}_d(00+)_{pv}&=&\frac{8\pi}{3}E_1^{Z_a^{\cal A}}M_1^\gamma v_T'\,,\\
S_0A^{00+}_{ed}(00+)_{pv}&=&\frac{8\pi}{3}\Big[
(E_1^\gamma+E_1^{Z_a^{\cal V}})M_1^\gamma v_T' 
+(C_0^\gamma C_0^{Z_v^{\cal A}}+C_2^\gamma C_2^{Z_v^{\cal A}})v_L
+M_1^\gamma M_1^{Z_v^{\cal A}} v_T\Big]\,.
\end{eqnarray}
The kinematic functions $v_\alpha^{(\prime)}$ ($\alpha\in|{L,T,LT,TT}$) 
reflecting the virtual photon density matrix are in detail
\begin{eqnarray}
\begin{array}{ll}
v_L=\beta^2\xi^2\,,\quad & v_T=\frac{1}{2}(2\zeta+\xi)\,,\\
 & \\
v_{LT}=2\beta\xi\,, & v_{TT}=-\frac{1}{2}\xi \,,\\
 & \\
v_{LT}'=\beta\sqrt{2\zeta}\,\xi\,, & v_T'=\sqrt{\zeta(\zeta+\xi)}\,,\\
\end{array}
\end{eqnarray}
with 
\begin{eqnarray}
\beta=\frac{|\vec q^{\,\mathrm{lab}}|}{|\vec q^{\,c}|}\,,\quad
\xi=-\frac{q_\nu^2}{|\vec q^{\,\mathrm{lab}}|^2}\,,\quad 
\zeta=\tan^2\frac{\theta_e^{\mathrm{lab}}}{2}\,,
\end{eqnarray}
where $\beta$ expresses the boost from the lab system to the frame in which
the multipoles are evaluated and $\vec q^{\,c}$ denotes the
three-momentum transfer in this frame. 

Collecting all terms one obtains
\begin{eqnarray}
\frac{d\sigma^{\gamma+Z}}{d\Omega_{k_2}^{\mathrm{lab}}}&=&\frac{4\pi}{3}
\,\sigma_{\mathrm{Mott}}\,\Big[((C_0^\gamma)^2+ (C_2^\gamma)^2)v_L
  +(M_1^\gamma)^2v_T 
+2E_1^{Z_a^{\cal A}}M_1^\gamma v_T'\nonumber\\&&
+2h\Big((E_1^\gamma+E_1^{Z_a^{\cal V}})M_1^\gamma v_T' 
+(C_0^\gamma C_0^{Z_v^{\cal A}}+C_2^\gamma C_2^{Z_v^{\cal A}})v_L
+M_1^\gamma M_1^{Z_v^{\cal A}} v_T\Big)\Big]\,,
\end{eqnarray}
where $C^c_L$, $E^c_L$, and $M^c_L$ are the charge and transverse electric 
and magnetic multipoles, respectively, defined in subsection IIB for
various hadronic currents $c\,(\gamma,\,Z_v,\, \mbox{or}\,Z_a)$ and
which are given explicitly below in section IV A.

In the Standard Model without radiative corrections and in the absence of the 
strangeness contribution to the neutral current, all multipole matrix
elements are proportional to the conventional electromagnetic ones.
According to the 
relations given in~\cite{ArS01}, we get for isoscalar transitions
\begin{eqnarray}
E_1^{Z_a^{\cal A}}&=& \widetilde G_aF^A_{E1}\,, 
\qquad E_1^{Z_a^{\cal V}}=\widetilde G_vF^A_{E1},\,\,\,\,\mbox{where}\,\,\,\,\,\,
 F^A_{E1}=\frac{\sqrt{E_d^\prime E_d}}{M_d}\,,
\langle 1||E_1\left(J^{Z_a}\right)|| 1\rangle\,,\\
M_1^{Z_v^{\cal A}}&=& g_v^d\widetilde G_a M_1^\gamma\,,
\qquad C_L^{Z_v^{\cal A}}=g_v^d\widetilde G_a C_L^\gamma,\,\,\,\,\mbox{where}\,\,
\,\,\,g_v^d=-2\sin^2\theta_W\,.
\end{eqnarray}
Furthermore, $C_L^\gamma$ ($L=0,2$) and $M_1^\gamma$ denote the usual
electromagnetic deuteron charge and magnetic multipoles, $F^A_{E1}$ its
electric multipole due to axial current and $E_1^\gamma$ the electric multipole due to a
parity violating admixture in the wave function. Thus, in the Standard Model
the cross section becomes 
\begin{eqnarray}
\frac{d\sigma^{\gamma+Z}}{d\Omega_{k_2}^{\mathrm{lab}}}&=&\frac{4\pi}{3}
\,\sigma_{\mathrm{Mott}}\,\Big[(1-h4\sin^2\theta_W\widetilde G_a)
\Big((C_0^\gamma)^2+ (C_2^\gamma)^2)v_L +(M_1^\gamma)^2v_T\Big)\nonumber\\&& 
+2\Big(hE_1^\gamma+(\widetilde G_a+h\widetilde G_v)F^A_{E1}\Big)M_1^\gamma
v_T'\Big]\,. 
\end{eqnarray}

\item{(ii)}
Vector recoil polarization:
\begin{eqnarray}
P_x\frac{d\sigma^{\gamma+Z}}{d\Omega_{k_2}^{\mathrm{lab}}}
&=&-\frac{1}{\sqrt{3}}\,\sigma_{\mathrm{Mott}}\,S_0\Big(A^{00+}_d(11+)_{pv}
+hA^{00+}_{ed}(11+)_{pc+pv}\Big)\,,\\
P_y\frac{d\sigma^{\gamma+Z}}{d\Omega_{k_2}^{\mathrm{lab}}}&=&0\,,\\
P_z\frac{d\sigma^{\gamma+Z}}{d\Omega_{k_2}^{\mathrm{lab}}}
&=&\sqrt{\frac{2}{3}}\,\sigma_{\mathrm{Mott}}\,S_0\Big(A^{00+}_d(10+)_{pv}
+hA^{00+}_{ed}(10+)_{pc+pv}\Big)\,,
\end{eqnarray}
where the asymmetry parameters $A^{00+}_d(11+)_{pv}$,
$A^{00+}_d(10+)_{pv}$, $A^{00+}_{ed}(11+)_{pc+pv}$, 
and $A^{00+}_{ed}(10+)_{pc+pv}$ are given by
\begin{eqnarray}
S_0A^{00+}_d(11+)_{pv}&=&\frac{4\pi}{3}\left(E_1^\gamma+E_1^{Z_a^{\cal V}}\right)
\left(2\sqrt{2}C_0^\gamma+C_2^\gamma\right)v_{LT}\nonumber\\
&&+\frac{2\sqrt{2}}{3}\left[\left(4C_0^{Z_v^{\cal A}}+\sqrt{2}C_2^{Z_v^{\cal A}}\right)
M_1^\gamma+\left(4C_0^\gamma+\sqrt{2}C_2^\gamma\right)M_1^{Z_v^{\cal A}}\right]v_{LT}^\prime\,,\\
S_0A^{00+}_{ed}(11+)_{pc}&=&\frac{2\sqrt{2}\pi}{3}\left(4C_0^\gamma+\sqrt{2}C_2^\gamma\right)
\,M_1^\gamma\,v_{LT}^\prime\,,\\
S_0A^{00+}_{ed}(11+)_{pv}&=&\frac{4\pi}{3}\left(2\sqrt{2}C_0^\gamma+C_2^\gamma\right)
E_1^{Z_a^{\cal A}}\,v_{LT}\,,\\
S_0A^{00+}_d(10+)_{pv}&=&2\sqrt{\frac{2}{3}}\pi\left[\left(E_1^\gamma+E_1^{Z_a^{\cal V}}\right)
M_1^\gamma\,v_T+M_1^\gamma\,M_1^{Z_v^{\cal A}}\,v_T^\prime\right]\,,\\
S_0A^{00+}_{ed}(10+)_{pc}&=&\sqrt{\frac{2}{3}}\pi\left(M_1^\gamma\right)^2\,v_T^\prime\\
S_0A^{00+}_{ed}(10+)_{pv}&=&2\sqrt{\frac{2}{3}}\pi\,E_1^{Z_a^{\cal A}}\,M_1^\gamma\,v_T\,.
\end{eqnarray}
Introducing for convenience for any of the current contributions
\begin{eqnarray}
C^c=4C^c_0+\sqrt{2}C^c_2\,,
\end{eqnarray}
the evaluation of the asymmetries yields
\begin{eqnarray}
P_x\frac{d\sigma^{\gamma+Z}}{d\Omega_{k_2}^{\mathrm{lab}}}
&=&-\frac{4\pi}{3\sqrt{6}}\,\sigma_{\mathrm{Mott}}\,
\Big[(E_1^\gamma+E_1^{Z_a^{\cal V}})C^\gamma v_{LT}
+(C^{Z_v^{\cal A}}M_1^\gamma +C^\gamma M_1^{Z_v^{\cal A}}) v_{LT}'\nonumber\\
&&+h(M_1^\gamma v_{LT}'+ E_1^{Z_a^{\cal A}}v_{LT})C^\gamma\Big]\,,\\ 
P_z\frac{d\sigma^{\gamma+Z}}{d\Omega_{k_2}^{\mathrm{lab}}}
&=&\frac{2\pi}{3}\,\sigma_{\mathrm{Mott}}\,\Big[
2(E_1^\gamma+E_1^{Z_a^{\cal V}}) v_{T}+2M_1^{Z_v^{\cal A}}v_T'
+h(M_1^\gamma v_T'+2E_1^{Z_a^{\cal V}}v_T)\Big]M_1^\gamma\,.
\end{eqnarray}

Using the lepton coupling from Eqs.\ (29) and (30) one has in the Standard Model
\begin{eqnarray}
P_x\frac{d\sigma^{\gamma+Z}}{d\Omega_{k_2}^{\mathrm{lab}}}
&=&-\frac{4\pi}{3\sqrt{6}}\,\sigma_{\mathrm{Mott}}\,
\Big[\Big(E_1^\gamma+(\widetilde G_v+h\widetilde G_a)F^A_{E1}\Big) v_{LT}
+(h-2\sin^2\theta_W\widetilde G_a)M_1^\gamma v_{LT}'\Big]C^\gamma\,,\\
P_z\frac{d\sigma^{\gamma+Z}}{d\Omega_{k_2}^{\mathrm{lab}}}
&=&\frac{2\pi}{3}\,\sigma_{\mathrm{Mott}}\,\Big[2\Big(E_1^\gamma
+(\widetilde G_v+h\widetilde G_a)F^A_{E1}\Big) v_{T}
+(h-4\sin^2\theta_W\widetilde G_a)M_1^\gamma v_T'\Big]M_1^\gamma\,.
\end{eqnarray}
We would like to point out, that the nonvanishing contribution to the vector
recoil polarization for unpolarized electrons is solely due to parity
violating contributions (see Eqs. (32)-(34)). 
\item{(ii)}
Tensor recoil polarization:\\
The nonvanishing components are
\begin{eqnarray}
P_{zz}\frac{d\sigma^{\gamma+Z}}{d\Omega_{k_2}^{\mathrm{lab}}}
&=&\frac{\sqrt{2}}{3}\,\sigma_{\mathrm{Mott}}\,S_0
\Big(A^{00+}_d(20+)_{pc+pv}+hA^{00+}_{ed}(20+)_{pv}\Big)\,,\\
P_{xx/yy}\frac{d\sigma^{\gamma+Z}}{d\Omega_{k_2}^{\mathrm{lab}}}
&=&\pm\frac{1}{2\sqrt{3}}
\,\sigma_{\mathrm{Mott}}\,S_0\Big(A^{00+}_d(22+)_{pc+pv}
+hA^{00+}_{ed}(22+)_{pv}\Big)
-\frac{1}{2}P_{zz}\frac{d\sigma^{\gamma+Z}}{d\Omega_{k_2}^{\mathrm{lab}}}\,,\\
P_{zx}\frac{d\sigma^{\gamma+Z}}{d\Omega_{k_2}^{\mathrm{lab}}}
&=&-\frac{1}{2\sqrt{3}}\,\sigma_{\mathrm{Mott}}\,S_0
\Big(A^{00+}_d(21+)_{pc+pv}+hA^{00+}_{ed}(21+)_{pv}\Big)\,,
\end{eqnarray}
where the parameters $A^{00+}_d(20+)_{pc+pv}$, $A^{00+}_d(21+)_{pc+pv}$, $A^{00+}_{ed}(20+)_{pv}$,
$A^{00+}_{ed}(21+)_{pv}$ and $A^{00+}_{ed}(22+)_{pv}$ are given by
\begin{eqnarray}
S_0A^{00+}_d(20+)_{pc}&=&-\frac{2\pi}{3}\left(4C_0^\gamma+\sqrt{2}C_2^\gamma\right)\,C_2^\gamma\,v_L
-\frac{\sqrt{2}\pi}{3}\left(M_1^\gamma\right)^2\,v_T\,,\\
S_0A^{00+}_d(20+)_{pv}&=&-\frac{2\sqrt{2}}{3}\pi\,E_1^{Z_a^{\cal A}}\,M_1^\gamma\,v_T^\prime\,,\\
S_0A^{00+}_d(21+)_{pc}&=&4\pi\,C_2^\gamma\,M_1^\gamma\,v_{LT}\,,\\
S_0A^{00+}_d(21+)_{pv}&=&4\pi\,C_2^\gamma\,E_1^{Z_a^{\cal A}}\,v_{LT}^\prime\,,\\
S_0A^{00+}_{ed}(20+)_{pv}&=&-\frac{4\pi}{3}\left(2\,C_0^{Z_v^{\cal A}}\,C_2^\gamma
+2\,C_0^\gamma\,C_2^{Z_v^{\cal A}}+\sqrt{2}\,C_2^\gamma\,C_2^{Z_v^{\cal A}}\right)v_L\nonumber\\
&&-\frac{2\sqrt{2}}{3}\pi\left[\left(M_1^\gamma\,M_1^{Z_v^{\cal A}}\right)v_T
+\left(E_1^\gamma+E_1^{Z_a^{\cal V}}\right)\,M_1^\gamma\,v_T^\prime\right]\,,\\
S_0A^{00+}_{ed}(21+)_{pv}&=&4\pi\left[\left(C_2^{Z_v^{\cal A}}\,M_1^\gamma+C_2^\gamma
\,M_1^{Z_v^{\cal A}}\right)\,v_{LT}+\left(E_1^\gamma+E_1^{Z_a^{\cal V}}\right)\,C_2^\gamma\,v_{LT}^\prime\right]\,,\\
S_0A^{00+}_{ed}(22+)_{pv}&=&\frac{4\pi}{\sqrt{3}}\,M_1^\gamma\,M_1^{Z_v^{\cal A}}\,v_{TT}\,.
\end{eqnarray}
Using these expressions, one gets
\begin{eqnarray}
P_{zz}\frac{d\sigma^{\gamma+Z}}{d\Omega_{k_2}^{\mathrm{lab}}}
&=&-\frac{2\pi}{9}\,\sigma_{\mathrm{Mott}}\,
\Big[\sqrt{2}C^\gamma C_2^\gamma v_L+(M_1^\gamma)^2v_T 
+2E_1^{Z_a^{\cal A}}M_1^\gamma v_T'
\nonumber\\&& 
-2h\Big(\sqrt{2}(2C_0^\gamma C_2^{Z_v^{\cal A}}+2C_0^{Z_v^{\cal A}}C_2^\gamma 
+\sqrt{2}C_2^\gamma C_2^{Z_v^{\cal A}})v_L
+(M_1^{Z_v^{\cal A}}v_T+(E_1^\gamma+E_1^{Z_a^{\cal
    V}})v_T')M_1^\gamma\Big)\Big]\,,\\ 
P_{xx/yy}\frac{d\sigma^{\gamma+Z}}{d\Omega_{k_2}^{\mathrm{lab}}}
&=&\pm\frac{\pi}{3}
\,\sigma_{\mathrm{Mott}}\,\Big[M_1^\gamma 
+h2M_1^{Z_v^{\cal A}}\Big]M_1^\gamma v_{TT}
-\frac{1}{2}P_{zz}\frac{d\sigma^{\gamma+Z}}{d\Omega_{k_2}^{\mathrm{lab}}}\,,\\
P_{zx}\frac{d\sigma^{\gamma+Z}}{d\Omega_{k_2}^{\mathrm{lab}}}
&=&-\frac{2\pi}{\sqrt{3}}\,\sigma_{\mathrm{Mott}}\,\Big[
C_2^\gamma M_1^\gamma v_{LT}+E_1^{Z_a^{\cal A}}C_2^\gamma v_{LT}'\nonumber\\&&
+h\Big((C_2^{Z_a^{\cal V}}M_1^\gamma +C_2^\gamma M_1^{Z_v^{\cal A}})v_{LT}
+(E_1^\gamma +E_1^{Z_a^{\cal V}})C_2^\gamma v_{LT}'\Big)\Big]\,.
\end{eqnarray}
Again, these expressions reduce in the Standard Model to 
\begin{eqnarray}
P_{zz}\frac{d\sigma^{\gamma+Z}}{d\Omega_{k_2}^{\mathrm{lab}}}
&=&-\frac{2\pi}{9}\,\sigma_{\mathrm{Mott}}\,
\Big[(1+h4\sin^2\theta_W\widetilde G_a)
\Big(\sqrt{2}C^\gamma C_2^\gamma v_L+(M_1^\gamma)^2v_T\Big)\nonumber\\&& 
+2\Big((\widetilde G_a-h\widetilde G_v)F^A_{E1} -hE_1^\gamma\Big)v_T'\Big]\,,\\
P_{xx/yy}\frac{d\sigma^{\gamma+Z}}{d\Omega_{k_2}^{\mathrm{lab}}}
&=&\pm\frac{\pi}{3}  
\,\sigma_{\mathrm{Mott}}\,(1-h4\sin^2\theta_W\widetilde G_a)
(M_1^\gamma)^2v_{TT}
-\frac{1}{2}P_{zz}\frac{d\sigma^{\gamma+Z}}{d\Omega_{k_2}^{\mathrm{lab}}}\,,\\
P_{zx}\frac{d\sigma^{\gamma+Z}}{d\Omega_{k_2}^{\mathrm{lab}}}
&=&-\frac{2\pi}{\sqrt{3}}\,\sigma_{\mathrm{Mott}}\,
\Big[(1-h4\sin^2\theta_W\widetilde G_a) 
M_1^\gamma v_{LT}+(E_1^\gamma+(\widetilde G_a
+h\widetilde G_v)F^A_{E1})v_{LT}'\Big]C_2^\gamma\,.
\end{eqnarray}

\end{description}

\section{Deuteron Wave Functions}
\subsection{Deuteron Parity Conserving Components}

The isospin singlet parity conserving component of the deuteron wave function 
$\psi_d^{pc}(\vec{r})$ is written as
\begin{equation}
\psi_d^{pc}(\vec{r})=\sum_{L=0,2}i^L\frac{u_L(r)}{r}~Y_{1L,1}^{m_d}(\hat{r}),
\end{equation}
where
\begin{equation}
Y_{L1,1}^{m_d}(\hat{r})=\sum_{m_L,m_S}\langle L m_L 1 m_S|1 m_d\rangle
Y_{LML}(\hat{r})\chi_{1m_S}, 
\end{equation}
with $\chi_{1m_S}$ denoting the deuteron spin wave function,
and the radial parts $u_L(r)$ are taken from the parameterization of
Machleidt et al.~\cite{MHC87}. for the Bonn OBEPQ model, given by
\begin{equation}
u_0(r)=\sum_{i=1}^{n_0}~C_i~e^{-m_ir},\qquad 
u_2(r)=\sum_{i=1}^{n_2}~D_i~e^{-m_ir}\left(1+\frac{3}{m_ir} 
+\frac{3}{(m_ir)^2}\right),
\end{equation}
with the normalization
\begin{equation}
\sum_{L=0,2}\int_0^\infty u_L^2(r)~dr~=~1.
\end{equation}
The constants $C_i$ and $m_i$ in Eq.\ (64) are listed in~\cite{MHC87}. 
\subsection{Deuteron Parity Violating Components}

The calculation of the small parity violating component in the deuteron due to
the weak parity violating hadronic potential $V^{pnc}$ is done in first order
perturbation theory as described in~\cite{KuA97} using the potential
of~\cite{DDH80} 

\begin{eqnarray}
 V^{pnc}(\vec r,\vec p\,) &=& i\frac{f_{\pi}g_{\pi{NN}}}{2\sqrt{2}M}
    ( \vec{\tau}_1 \times \vec{\tau}_2)_z (\vec{\sigma}_1+
    \vec{\sigma}_2)\cdot 
 \Big[ \vec{p} ,f_{\pi}(r) \Big]  \nonumber \\
 & & -\frac{g_{\rho}}{M} \Big(h_{\rho}^0 \,\vec{\tau}_1\cdot \vec{\tau}_2+
     \frac{h_{\rho}^1}{2} ( \vec{\tau}_1 + \vec{\tau}_2)_z+
     \frac{h_{\rho}^2} {2 \sqrt{6}} (3 \tau_{1,z} \tau_{2,z} -\vec{\tau}_1\cdot 
     \vec{\tau}_2) \Big) \nonumber \\
 & & \times \Big( (\vec{\sigma}_1-\vec{\sigma}_2) \cdot 
     \Big\{ \vec{p},f_{\rho}(r) \Big\} 
   + i(1+\chi_v) \ (\vec{\sigma}_1 \times \vec{\sigma}_2)
    \cdot \Big[ \vec{p},f_{\rho}(r) \Big] \,\Big) 
\nonumber \\
 & & - \frac{g_{\omega}}{M} \Big( h_{\omega}^0 + \frac{h_{\omega}^1}{2} 
     ( \vec{\tau}_1 + \vec{\tau}_2)_z\, \Big) 
\nonumber \\
 & &  \times 
\Big( (\vec{\sigma}_1-\vec{\sigma}_2) \cdot 
    \Big\{ \vec{p},f_{\omega}(r) \Big\} 
    + i(1+\chi_s) \ (\vec{\sigma}_1 \times \vec{\sigma}_2)
    \cdot \Big[ \vec{p},f_{\omega}(r) \Big] \,\Big)
\nonumber \\
 &  & -\frac{1}{2M}\,
      ( \vec{\tau}_1 - \vec{\tau}_2)_z 
      (\vec{\sigma}_1+\vec{\sigma}_2) \cdot 
      \Big\{ \vec{p},g_{\omega} h_{\omega}^1f_{\omega}(r) 
      - g_{\rho} h_{\rho}^1f_{\rho}(r) \Big\}\nonumber \\
 & & -i\frac{g_{\rho} h_{\rho}^{\prime 1}}{2M}( \vec{\tau}_1 
     \times \vec{\tau}_2)_z (\vec{\sigma}_1+\vec{\sigma}_2)\cdot 
     \Big[ \vec{p},f_{\rho}(r) \Big],
\label{pv_pot}
\end{eqnarray}
with the usual Yukawa function 
\begin{eqnarray}
f_{\xi}(r) =  \frac{e^{-m_{\xi}r}}{4\pi r} \,, \quad \mbox{for} 
\quad \xi=\pi,\rho,\omega\,.
\end{eqnarray}
$M$ denotes the nucleon mass and $\vec p= \frac{1}{2}(\vec p_1
-\vec p_2)$. Values for the weak coupling constants for various models
of parity violating hadronic potentials are listed in
Table I. 
\begin{table}[h]
\renewcommand{\arraystretch}{1.2}
\caption{\label{coupling_constants}
Weak coupling constants for various models of 
parity violating $NN$ interaction $V^{pnc}$ in 
units of $g_\pi=3.8\times 10^{-8}$.
}

\begin{tabular}{|c|c|c|c|c|c|c|c|}\hline
Coupling&$f_\pi$ & $h_\rho^0$ & $h_\rho^1$ & $h_\rho^2$ & 
$h_\omega ^0$ & $h_\omega ^1$ &$h_\rho^{\prime 1}$\\
\hline\hline
DDH~\cite{DDH80}&$12$ & $-30$ & $-0.5$ & $-25$ & $-5$ & $-3$&0\\
DZ~\cite{DZ86}&$+3$&$-22$&$+1$&$-18$&$-10$&$-6$&0\\
FCDH~\cite{FCDH91}&$+7$&$-10$&$-1$&$-18$&$-13$&$-6$&0\\
\hline
\end{tabular}

\end{table}\\
The small parity violating component in the deuteron wave function
$\psi_d^{pnc}$ is given in first order perturbation theory by
\begin{equation}
|~\psi_d^{pnc}~\rangle=-\frac{1}{H_{pc}^{(0)}-E}~
\left(H_{pnc}^{(1)}-\left(E-E_B\right)\right)|~\psi_d^{pc}~\rangle \,,
\end{equation}
where $|~\psi_d^{pc}~\rangle$ is the unperturbed parity conserving
deuteron wave function, $H_{pc}^{(0)}$
and $H_{pnc}^{(1)}$ are the parity conserving and violating strong
Hamiltonians, respectively, $E_B$ is the unperturbed deuteron binding
energy (=--2.2246 MeV) and $E$ the eigenvalue of the perturbed wave
function. In the present calculation, the propagator was approximated
by the free Greens function. Applying the potential of Eq.\ (66) to the
unperturbed wave function yields for the parity violating admixture 

\begin{eqnarray}
 \psi_d^{pnc}({\bf p})&=& \frac{i}{p}\Big(
\tilde{u}_{11}(p)\,\langle \hat p|10;(11)1m_d\rangle
+\tilde{u}_{10}(p)\,\langle \hat p|00;(10)1m_d\rangle\Big)\,,
\end{eqnarray}

where $\tilde{u}_{1S}(p)$ denotes the radial part and 
$\langle \hat p|Tm_T;(1S)JM\rangle$ the isospin, orbital and spin
angular momentum parts of the parity violating p-wave (L=1) components
of the deuteron. The two contributions $u_{11}(p)$ and $u_{10}(p)$
correspond to isovector $^3P_1$ and isoscalar $^1P_1$ states,
respectively. In detail one finds  
\begin{equation}
\tilde{u}_{1S}(p)=\frac{u_{1S}(p)}{p}=\frac{1}{E_B-\frac{p^2}{M}}
\int dr~j_1(pr)~f_{1S}(r),~~~~~E_B < 0\,,
\end{equation}
where for the triplet state $\tilde{u}_{11}(p)$ the function
$f_{11}(r)$ is given by 
\begin{eqnarray}
f_{11}(r)&=&-\frac{1}{\pi
  M\sqrt{3\pi}}\sum_{L=0,2}(\sqrt{2})^{-L/2}\left\{\frac{f_\pi g_{\pi
      NN}}{\sqrt{2}}~e^{-m_\pi
    r}\left(m_\pi+\frac{1}{r}\right)u_L(r)\right.\nonumber\\  
&&\left.-g_\rho h_\rho^{\prime 1}~e^{-m_\rho r}\left(m_\rho+\frac{1}{r}\right)u_L(r)
-g_\omega h_\omega^1~e^{-m_\omega
  r}\left[\left(m_\omega+(-)^{L/2}\frac{3}{r}\right)u_l(r)-2u_L^\prime(r)
\right]\right.\nonumber\\  
&&\left.+g_\rho h_\rho^1~e^{-m_\rho
    r}\left[\left(m_\rho+(-)^{L/2}\frac{3}{r}\right)
u_l(r)-2u_L^\prime(r)\right]\right\}\,, 
\end{eqnarray}
and for the singlet state $\tilde{u}_{10}(p)$
\begin{eqnarray}
f_{10}(r)&=&-\frac{1}{\pi M\sqrt{6\pi}}\sum_{L=0,2}(-\sqrt{2})^{-L/2}\left\{-3g_\rho h_\rho^0~e^{-m_\rho r}\left(\left[\left(m_\rho+\frac{1}{r}\right)\chi_v+\frac{(-2)^{1+L/2}}{r}\right]u_L(r)+2u_L^\prime(r)\right)\right.\nonumber\\
&&\left.+g_\omega h_\omega^0~e^{-m_\omega r}\left(\left[\left(m_\rho+\frac{1}{r}\right)\chi_S+\frac{(-2)^{1+L/2}}{r}\right]u_L(r)+2u_L^\prime(r)\right)\right\}\,.
\end{eqnarray}
In configration space, the parity violating components $u_{1S}(r)$ are
obtained by taking the Fourier transform of the momentum space wave
functions $u_{1S}(p)$ given in Eq.~(70). Explicitly they are
written as~\cite{ADW71} 
\begin{eqnarray}
\frac{u_{1S}(r)}{r}&=&\sqrt{\frac{2}{\pi}}\int_0^\infty~dp~p^2~\frac{u_{1S}(p)}{p}~j_1(pr)\nonumber\\
&=&-\sqrt{\frac{2}{\pi}}M\int_0^\infty~dr^\prime~G_1(r,r^\prime)~f_{1S}(r),
\end{eqnarray}
where
\begin{equation}
G_1(r,r^\prime)=\int_0^\infty dp~p^2~\frac{j_1(pr)~j_1(pr^\prime)}{p^2+\epsilon^2}=\frac{1}{rr^\prime}H_1(r,r^\prime,\epsilon),
\end{equation}
with
\begin{equation}
\epsilon^2=-E_BM~~~\mbox{and}~~~H_1(r,r^\prime,t)=\epsilon\,k_1(\epsilon
r_{>})i_1(\epsilon r_{<}),
\end{equation}
where $k_1$ and $i_1$ are the modified Bessel and Hankel
functions. For a given parity conserving wave function of the
deuteron, the p-wave components are numerically quite sensitive to the
parameters of the weak nucleon-nucleon potential, especially to
$f_\pi$. These p-wave components are shown in Fig.~1 for various
potential models whose parameters values are given in Table~I.  

\section{Electroweak Currents and Multipoles}
\subsection{Multipoles}
We see from section II, that in elastic deuteron scattering with
unpolarized deuteron targets, the various observables like
differential cross section $d\sigma/d\Omega$, vector and tensor
polarizations of the recoil deuteron, i.e. $P_i~(i=x,y,z)$ and
$P_{ij}~(i,j=x,y,z)$, are given in terms of the quantities
$A_d^{00+}(IM)$ and $A_{ed}^{00+}(IM)$, with $I=0,1,2$ and $0<M<I$
for unpolarized and polarized electron scattering. These quantities
are defined in terms of
multipoles $C_0^c$, $C_2^c$, $E_1^c$, and $M_1^c$ which are reduced
matrix elements of the general multipole operators (see Eq.\ (14)) for
a given current $c$, which may be the vector current $(J^\gamma)$ of
the electromagnetic interaction due to one photon exchange or the weak
neutral vector and axial currents $(J^{Z_v}~\mbox{and}~J^{Z_a})$ 
due to $Z$ exchange. It should be noted that in the presence of 
parity violation the nonvanishing multipole $E_1^c$  arises from the
weak axial vector current $(J^{Z_a})$ and also from the
electromagnetic current $(J^\gamma)$ when the parity violating
component is included in the wave function of the deuteron. The
general multipole operators are defined by~\cite{WAL}. 
\begin{eqnarray}
C_{LM}^c&=&\int d\vec{x}\left[j_L(qx)~\vec{Y}_{LM}(\hat x)\right]{J_0}^c(\vec{x})\,,\\
E_{LM}^c&=&\frac{1}{q}\int d\vec{x}~\vec{\nabla}\times\left[j_L(qx)~\vec{Y}_{LL1}^M
(\hat x)\right]\cdot\vec{J}^c(\vec{x})\,,\\
M_{LM}^c&=&\int d\vec{x}\left[j_L(qx)~\vec{Y}_{LL1}^M(\hat
  x)\right]\cdot\vec{J}^c(\vec{x})\,,
\end{eqnarray}
where the currents $J_\mu^c\,(c=\gamma, Z_v, Z_a)$ are given by their
single nucleon matrix elements in terms of the weak and
electromagnetic form factors of the nucleons as follows
\begin{eqnarray}
\langle p^\prime|J_\mu^{\gamma,p(n)}|p\rangle&=&\bar{u}(p^\prime)\left[F_1^{\gamma,p(n)}(Q^2)
\gamma_\mu+iF_2^{\gamma,p(n)}(Q^2)\frac{\sigma_{\mu\nu}}{2M}q^\nu\right]u(p)\,,\\
\langle p^\prime|J_\mu^{Z,p(n)}|p\rangle&=&\langle
p^\prime|J_\mu^{Z_v,p(n)}+J_\mu^{Z_a,p(n)}|p\rangle\,,
\end{eqnarray}
with
\begin{eqnarray}
\langle p^\prime|J_\mu^{Z_v,p(n)}|p\rangle&=&\bar{u}(p^\prime)\left[F_1^{Z,p(n)}(Q^2)
\gamma_\mu+iF_2^{Z,p(n)}(Q^2)\sigma_{\mu\nu}\frac{q^\nu}{2M}\right]u(p)\,,\\
\langle p^\prime|J_\mu^{Z_a,p(n)}|p\rangle&=&\bar{u}(p^\prime)\left[G_A^{Z,p(n)}(Q^2)\gamma_\mu\gamma_5
\right]u(p)\,,
\end{eqnarray}
where $F_1^{\gamma,Z}(Q^2)$ and $F_2^{\gamma,Z}(Q^2)$ ($Q^2=-q_\mu^2$)
are the Dirac form factors of the electromagnetic and weak neutral vector
currents which are related to the Sachs form factors
$G_E^{\gamma,Z}(Q^2)$ and $G_M^{\gamma,Z}(Q^2)$ as follows 
\begin{eqnarray}
G_E^{\gamma,Z}(Q^2)&=&F_1^{\gamma,Z}(Q^2)-\tau\,F_2^{\gamma,Z}(Q^2)\,,\\
G_M^{\gamma,Z}(Q^2)&=&F_1^{\gamma,Z}(Q^2)+F_2^{\gamma,Z}(Q^2),\,\,\,\tau=\frac{Q^2}{4M}\,.
\end{eqnarray}
The weak neutral current form factors $G_{E,M}^Z(Q^2)$ are defined in the
standard model~\cite{ABM02} by
\begin{equation}
G_{E,M}^{Z,p(n)}(Q^2)=\frac{1}{2}\left(1-4\sin^2\theta_W\right)G_{E,M}^{p(n)}(Q^2)-\frac{1}{2}
G_{E,M}^{n(p)}(Q^2)\,,
\end{equation}
where $\theta_W$ is the weak mixing angle in the standard model.
The axial vector form factors are given by
\begin{equation}
G_A^{Z,p(n)}(Q^2)=-\frac{\tau_3}{2}G_A^{Z}(Q^2),\,\,\,\,\,\tau_3=+1(-1)\,\,\mbox{for p(n)}.
\end{equation}
In the presence of nonzero strangeness of the nucleon, these form
factors are modified as
\begin{eqnarray}
G_{E,M}^{Z,p(n)}(Q^2)&\rightarrow&G_{E,M}^{Z,p(n)}(Q^2)-\frac{1}{2}G_{E,M}^s(Q^2)\,,\\
G_A^{Z,p(n)}(Q^2)&\rightarrow&G_A^{Z,p(n)}(Q^2)+\frac{1}{2}G_A^S(Q^2)\,.
\end{eqnarray}
The matrix elements of the multipole operators between the initial and
final deuteron state are calculated using the nonrelativistic limit of
the current matrix elements for the vector and axial currents given by:
\begin{eqnarray}
\left\{J_0,\,\vec{J}_{\gamma}\right\}&=&\chi_{s_f}^\dagger\left\{G_E^{\gamma}(Q^2),\,\frac{1}{2M}\left[G_E^{\gamma}(Q^2)
\left(\vec{p}_f+\vec{p}_i\right)+iG_M^{\gamma}(Q^2)\left(\vec{\sigma}\times\vec{q}\,\right)
\right]\right\}\chi_{s_i}\,,\\
\left\{J_0,\,\vec{J}{_{Z_v}}\right\}&=&\chi_{s_f}^\dagger\left\{G_E^{Z}(Q^2),\,\frac{1}{2M}\left[G_E^{Z}(Q^2)
\left(\vec{p}_f+\vec{p}_i\right)+iG_M^{Z}(Q^2)\left(\vec{\sigma}\times\vec{q}\,\right)
\right]\right\}\chi_{s_i}\,,\\
\left\{J_0,\,\vec{J}_{{Z_a}}\right\}&=&\chi_{s_f}^\dagger\left\{\frac{G_A^{Z}(Q^2)}{2M}\vec{\sigma}\cdot
\left(\vec{p}_i+\vec{p}_f\right),\,-G_A^{Z}(Q^2)\vec{\sigma}\right\}\chi_{s_i}\,.
\end{eqnarray}
\subsection{Electroweak Form Factors}
\subsubsection{Electromagnetic and Weak Form Factors}
The electromagnetic form factors $G_{E,M}^{\gamma,p(n)}(Q^2)$ defined in Eqs.\ (83) and (84) are
generally parameterized in dipole forms given by
\begin{eqnarray}
G_E^p(Q^2)&=&\left[1+\frac{Q^2}{M_V^2}\right]^{-2},~~~~~~~~G_E^n(Q^2)=\mu_n\xi_n\frac{Q^2}{M^2}
G_E^p(Q^2)\,,\\
G_M^p(Q^2)&=&(1+\mu_p)G_E^p(Q^2),~~~~~G_M^n(Q^2)=\mu_nG_E^p(Q^2)\,,
\end{eqnarray}
with
\begin{eqnarray}
\xi_n&=&\left[1-\lambda_n\left(\frac{Q^2}{4M^2}\right)\right]^{-1},\quad\mu_p=1.792847,
\quad\mu_n=-1.913043,\quad\lambda_n=5.6\,. 
\end{eqnarray}
The axial vector form factor $G_A^{Z,p(n)}(Q^2)$ is also parameterized in dipole form as
\begin{eqnarray}
G_A^{Z,p(n)}(Q^2)&=&G_A^{Z,p(n)}(0)\left[1+\frac{Q^2}{M_A^2}\right]^{-2}\mbox{ with }
\quad G_A^{Z,p(n)}(0)=-\frac{1}{2}\left(+\frac{1}{2}\right)1.262\,.
\end{eqnarray}
The numerical value of the vector dipole mass $M_V=$ 0.84 GeV is taken
from experimental data on electron proton scattering, and the axial
dipole mass $M_A=$ 1.026 GeV from neutrino scattering from proton and
deuteron~\cite{SKS02}.
\subsubsection{Strangeness Form Factors}
A number of theoretical models have been used to establish the magnitude of the strange quark form factor 
and its $Q^2$ dependence. They use either a pole with simple vector dominance for the vector form
factors or a meson cloud model which considers kaon loops and other higher strange resonance loop
contributions to calculate the strangeness form factors. A review of many of these models is given in~\cite{BH01}.
Recently, chiral perturbation theory~\cite{HPRZ03} and 
lattice QCD~\cite{LT00,LWW03} have also been applied for their calculation. 
In the earlier theoretical work a dipole form has been used.
However, in the present calculation we have used for the vector
strangeness form factor the recently determined form factors from a
global analysis of presently available data on parity violating
electron scattering by Liu et al.~\cite{LiMR07}, i.e. with $Q^2$ in (GeV/c)$^2$
\begin{eqnarray}
G_E^s(Q^2)&=&G_E^s\,\frac{Q^2}{0.1},~~~~~~~~~~~~~~~~~~~G_E^s=-0.014\,\\
G_M^s(Q^2)&=&G_M^s+\mu_s^\prime\left(Q^2-0.1\right),\,\,\,G_M^s=0.28,\,\,\,\mu_s^\prime=-0.1 \mbox{(GeV/c)}^{-2}\,.
\end{eqnarray}
For the axial vector form factor $G_A^s(Q^2)$ we use a dipole form
\begin{equation}
G_A^s(Q^2)=g_A^s(0)\left[1+\frac{Q^2}{M_A^2}\right]^{-2}\,,
\end{equation}
with $g_A^s(0)=\Delta s=-0.19$ and $M_A=$ 1.026 GeV, where $\Delta s$
is the spin contribution of the strange quarks and antiquarks
determined experimentally from deep inelastic scattering of
electrons. 
\subsubsection{Radiative Corrections and Anapole Moments}
Higher order radiative corrections to the weak neutral current
couplings of the nucleon have been calculated in the standard
model~\cite{Pa04}-\cite{MH90}. The corrections to the vector form
factors are dominated by single quark transition but their effect
is found to be small as they are multiplied by the factor
$(1-4\sin^2\theta_W)$. In the presence of radiative corrections, the
weak vector form factors are written as 
\begin{equation}
G_{E,M}^{Z,p(n)}(Q^2)=\frac{1}{2}\left(1-4\sin^2\theta_W\right)\left(1+R_V^p\right)G_{E,M}^p
(Q^2)-\frac{1}{2}\left(1+R_V^n\right)G_{E,M}^n(Q^2)-\frac{1}{2}G_{E,M}^s(Q^2)\,,
\end{equation}
where $R_V^p$ and $R_V^n$ are the radiative corrections for proton and
neutron. In the case of the axial vector, the single quark and two quark
transitions are both important and lead to appreciable corrections. The
axial vector form factors in the presence of radiative correction are
written as 
\begin{equation}
G_A^Z(Q^2)=-\frac{1}{2}\tau_3\left(1+R_A^1\right)G_A-\frac{1}{2}R_A^0+\frac{1}{2}G_A^s\,, 
\end{equation}
where $R_A^1$ and $R_A^0$ are the radiative corrections in the
isovector and isoscalar channels. In addition to single quark
transition the two quark transition induce the following axial anapole
term in the matrix element of the electromagnetic current
\begin{equation}
\langle p^\prime|J_\mu^\gamma|p\rangle=\frac{Q^2}{M^2}\,\bar{u}(p^\prime)
\left[\gamma_\mu-\frac{\not{\!q}q_\mu}{q^2}\right]\gamma_5\,u(p)\left[a_S(Q^2)+a_V(Q^2)\tau_3\right]\,,
\end{equation}
which in leading order of the nonrelativistic limit is given as
\begin{equation}
\vec{J}=\frac{Q^2}{M^2}\,\chi_{s_f}^\dagger\left[\vec{\sigma}-\vec{\sigma}\cdot\hat{q}\,\hat{q}
\right]\chi_{s_i}\left[a_S(Q^2)+a_V(Q^2)\tau_3\right]\,
\end{equation}
with $\hat{q}$ as unit vector along $\vec{q}$.
It is equivalent to an axial coupling contributing to the axial form factor.
The coefficients $a_S(Q^2)$ and $a_V(Q^2)$ are calculated using pion
loop contributions in terms of  
the parity violating and parity conserving $\pi$NN couplings $f_{\pi NN}$ and $g_{\pi NN}$, and 
are given as~\cite{ZhPHR00}
\begin{equation}
a_{S,V}(0)=\frac{f_{\pi NN}\,g_{\pi
    NN}}{4\sqrt{2}\pi^2}\,\alpha_{S,V}(0)\,,\mbox{ with }\,\alpha_S(0)=1.6,
\,\,\alpha_V(0)=0.4\,.
\end{equation}
The contributions of these terms to the radiative corrections $R_A^1$ and $R_A^0$ are 
found to be small~\cite{ZhPHR00,ABM02}. The updated values of the radiative 
corrections, taken from ref.~\cite{BePS05}, are given in Table II.
\begin{table}[h]
\renewcommand{\arraystretch}{1.2}
\caption{Values of radiative corrections to weak neutral current couplings.}
\begin{tabular}{|c|c|c|}\hline
Correction & Isoscalar & Isovector \\ \hline\hline 
$R_V$ & -0.0113 & -0.017$\pm$0.002\\
$R_A$ & 0.06$\pm$0.14 & -0.23$\pm$0.24\\
\hline
\end{tabular}
\end{table}\\
\section{Results and Discussions}
The differential cross sections $d\sigma/d\Omega$, the vector and
tensor polarizations $P_i~(i=x,y,z)$ and $P_{ij}~(i,j=x,y,z)$,
respectively, are described 
in terms of various asymmetry parameters $A_d^{00+}(IM)$ and
$A_{ed}^{00+}(IM)$. They get contributions 
from the parity conserving as well as from the parity and time
reversal (T) violating pieces. In Table~III, we give a classification
of parity conserving and parity violating contributions to
$A_d^{00+}(IM)$, and $A_{ed}^{00+}(IM)$ 
for various values of $I$ and $M$. 
\begin{table}[h]
\caption{\label{coupling_constant}
Schematic survey of nonvanishing scalar asymmetries $A_d^{00+}(IM)$
and $A_{ed}^{00+}(IM)$ marked by ``$\surd$''.}
\begin{tabular}{|c|c|c|c|c|c|c|c|c|c|c|}\hline
Type & Current & 00+ & 10+ & 11+ & 11- & 20+ & 21+ & 21- & 22+ & 22- \\ \hline\hline
 & PT-conserving & $\surd$ & & & & $\surd$& $\surd$& & $\surd$&\\
$A_d^{00+}(IM)$ & P-violating & $\surd$& $\surd$& $\surd$& & $\surd$& $\surd$& & &\\
\hline\hline
 & PT-conserving & & $\surd$& $\surd$& & & & & &\\
$A_{ed}^{00+}(IM)$ & P-violating & $\surd$& $\surd$& $\surd$& &
$\surd$& $\surd$& & $\surd$&\\ 
\hline
\end{tabular}
\end{table}\\
We see that $A_d^{00+}(00+)$, $A_d^{00+}(20+)$, $A_d^{00+}(21+)$,
$A_{ed}^{00+}(10+)$, and $A_{ed}^{00+}(11+)$ receive contributions from
both parity violating as well as parity conserving pieces. It is
therefore not possible to study any parity violating effects through
observation of any of these asymmetry parameters as they will be
completely swamped by the parity conserving contributions. The only
way to study parity violating effects in e-d scattering with an
unpolarized deuteron target will be through the asymmetry parameters
$A_d^{00+}(10+)$ and $A_d^{00+}(11+)$ with unpolarized electron
scattering and through $A_{ed}^{00+}(00+)$, $A_{ed}^{00+}(20+)$,
$A_{ed}^{00+}(21+)$, and $A_{ed}^{00+}(22+)$ with polarized electron
scattering. These are related to the vector polarization of the
deuteron $P_x$ and $P_z$, electron beam asymmetry ${\cal A}$ and
various tensor polarization asymmetries ${\cal A}_{zz}$, ${\cal
  A}_{xx/yy}$ and ${\cal A}_{zx}$. Observational quantities are
defined in terms of the well known deuteron charge $G_C(Q^2)$,
magnetic moment $G_M(Q^2)$, and quadrupole $G_Q(Q^2)$ form factors as
well as the new axial vector form factor $G_A(Q^2)$ and electric form
factor $G_E(Q^2)$ which are defined in terms of various multipoles 
as follows:
\begin{eqnarray}
G_C(Q^2)&=&\sqrt{\frac{4\pi}{3}}\,\frac{\beta}{1+\eta}\,C_0\,,\\
G_Q(Q^2)&=&\sqrt{\frac{3\pi}{2}}\,\frac{\beta}{\eta\left(1+\eta\right)}\,C_2\,,\\
G_{E/M}(Q^2)&=&\sqrt{\frac{\pi}{\eta\left(1+\eta\right)}}\,\left(E/M\right)_L\,,\\
G_A(Q^2)&=&\sqrt{\frac{\pi}{\eta\left(1+\eta\right)}}\,F_{E1}^A\,.
\end{eqnarray}
In the following we present results for these asymmetries and discuss
the possibility to experimentally determine them. 
\subsection{e-d Scattering with Unpolarized Electrons}\label{unpolscat}
In the case of elastic electron deuteron scattering with unpolarized
electrons $(h=0)$ (see Eqs.\ (46)-(48)) and unpolarized deuterons, the
differential cross sections (Eq.\ (22)) and various 
tensor polarization components are dominated by the parity conserving
contributions and there is no hope of observing any parity violating
effects in these observables. However, the vector polarization $P_x$
and $P_z$ depend solely upon the parity violating contributions
$A_d^{00+}(10+)$ and $A_d^{00+}(11+)$  (Eqs.\ (32)-(34)). Neglecting
the parity violating contributions to the total cross section, we obtain 
\begin{eqnarray}
S_0P_z&=&\sqrt{\frac{2}{3}}S_0A_d^{00+}(10+)\nonumber\\
&=&\frac{2}{3}\eta\left[1+2\left(1+\eta\right)\tan^2\frac{\theta}{2}\right]
\left[G_{E}+\tilde{G_v}G_A\right]G_M
+\frac{4}{3}\sec\frac{\theta}{2}\tan\frac{\theta}{2}\,\eta\nonumber\\
&&\times\sqrt{\left(1+\eta\right)\left(1+\eta\sin^2
\frac{\theta}{2}\right)}g_v^d\tilde{G_a}G_M^2\,,\\ 
S_0P_x&=&-\frac{1}{\sqrt{3}}S_0A_d^{00+}(11+)\nonumber\\
&=&-\frac{4}{3}\sec\frac{\theta}{2}\sqrt{\eta
\left(1+\eta\sin^2\frac{\theta}{2}\right)}
\left[G_{E}+\tilde{G_v}G_A\right]\left[G_C+\frac{\eta}{3}G_Q\right]\nonumber\\
&&-\frac{8}{3}\tan\frac{\theta}{2}\sqrt{\eta\left(1+\eta\right)}g_v^d\tilde{G_a}
\left[G_C+\frac{\eta}{3}G_Q\right]G_M\,,
\end{eqnarray}
where $A_d^{00+}(11+)$ and $A_d^{00+}(10+)$ are given in Eqs.\ (35) and (38). 
Both polarization observables $P_x$ and $P_z$ depend upon the
contribution containing $G_{E}+\tilde{G_v}G_A$ in addition to the
Standard Model contribution proportional to $\tilde{G_a}$ which is
dominant. The contribution of $G_{E}$ depends upon the p-wave
component of the deuteron wave function while $G_A$ depends upon
the isoscalar axial vector form factor, which may be due to an isoscalar
strangeness component of the nucleon or an anapole moment form factor,
which both are expected to be small. We have considered these effects
including a strangeness component in the magnetic moment form factor 
as well, which are parameterized as given in Eqs. (96), (97) and (101)
with $g_A^s$ = -0.19. 

In Fig.2, we present the numerical values of $P_x$ and $P_z$ as a
function of $Q^2$ for various electron energies, corresponding to
future studies of parity violating effects in electron deuteron
scattering to be done at MAINZ~\cite{Ba07} and JLAB~\cite{Ro06}. We
see that the effect of parity violation via P-odd admixtures is 
very small as compared to the contribution from $\gamma$-Z
interference. In order to see the effect of the strangeness form
factors in magnetic as well as axial vector form factors, we have
considered three cases: (i) $G_M^S\neq$ 0, $G_A^S$ = 0, and $G_E^S$ =
0, (ii) $G_M^S$ = 0, $G_A^S\neq$ 0, and $G_E^S$ = 0, and
(iii) $G_M^S\neq$ 0, $G_A^S\neq$ 0, and $G_E^S$ = 0, and show the
effect in Figs.~3 and 4 as a function of $Q^2$, and as a function of
$\theta$, for fixed $Q^2$, at various electron energies. For this
purpose, we have neglected the $G_E$ contribution to $P_x$ and $P_z$,
which is quite small. One readily notes that, if presently suggested values of
$G_M^S$ and $G_A^S$ are taken, then the effect of a strangeness
component in the axial vector and magnetic moments are opposite in
sign and tend to cancel each other. Moreover, the magnetic effect of
a strangeness component in the magnetic moment is larger than the
effect of including a strangeness component in the axial vector form
factor if presently suggested values for their magnitudes are
used. The parity violating effects in elastic deuteron scattering have
been considered before by many authors~\cite{Po81}-\cite{HwH80}. But 
none of them has calculated the vector polarization of the recoil
deuteron, although it is discussed qualitatively by Ramachandran and
Singh~\cite{RaS78}. 

\subsection{e-d Scattering with Polarized Electrons}
In the case of elastic electron scattering with polarized electrons
$(h\neq 0)$ on unpolarized deuterons, the vector polarization
components of the recoil deuteron are dominated by the parity
conserving contributions $A_{ed}^{00+}(11+)$ and $A_{ed}^{00+}(10+)$,
and thus no useful information on parity violation can be obtained
from observing $P_x$ and $P_z$. In this case the purely parity
violating contributions $A_{ed}^{00+}(00+)$, $A_{ed}^{00+}(20+)$, 
$A_{ed}^{00+}(21+)$, and $A_{ed}^{00+}(22+)$ (see Table III) can be
determined through appropriately defined asymmetries of 
the differential cross section and the recoil tensor polarizations. For
example, defining the beam asymmetry as 
\begin{equation}
{\cal A}=\frac{1}{h}\frac{\left[\left(\frac{d\sigma}{d\Omega}\right)_+
-\left(\frac{d\sigma}{d\Omega}\right)_-\right]}
{\left[\left(\frac{d\sigma}{d\Omega}\right)_++
\left(\frac{d\sigma}{d\Omega}\right)_-\right]}\,,
\end{equation}
where $\left(\frac{d\sigma}{d\Omega}\right)_{\pm}$ denotes the
differential cross section with right handed $(h=+1)$ and left handed
$(h=-1)$ polarized electrons, then one obtains ${\cal
  A}=A_{ed}^{00+}(00+)$. It is given, separating it into ${\cal A}_Z$
and ${\cal A}_\gamma$, by 
\begin{equation}
{\cal A}={\cal A}_Z+{\cal  A}_\gamma
=A_{ed}^{00+}(00+)=2g_v^d\tilde{G_a}+\frac{8}{3S_0}\sec\frac{\theta}{2} 
\tan\frac{\theta}{2}\eta\,\sqrt{\left(1+\eta\right)\left(1+\eta\sin^2
\frac{\theta}{2}\right)} \left(G_E+\tilde{G_v}G_A\right)G_M\,.
\end{equation}
In Figs.~5 and 6, we show this asymmetry 
as a function of $Q^2$ as well as a function of the scattering angle
$\theta$, for fixed $Q^2$, at various energies available for
experiments at MAINZ~\cite{Ba07} and JLAB~\cite{Ro06}. We have
also studied the effect of a strangeness form factor on these
asymmetries for various cases discussed in subsection \ref{unpolscat}
using the functional form of the strangeness form factor given in
Eqs. (96), (97) and (101). For this purpose, we have again neglected the
contribution of $G_E$. 
We see from Eq.~(111) that apart from the dominant contribution 
proportional to $\tilde{G_a}$, the additional contribution 
depends upon $G_{E}+\tilde{G_v}G_A$, the same combination of weak form 
factors as encountered in $P_x$ and $P_z$, but with a different angular
dependence.  

We see that the dominant contribution to ${\cal A}$ comes from the
$\gamma$-Z interference whereas the contribution of $G_E$, coming
from the parity violating effect in the wave function, is small. This
is similar to the results found in quasi-elastic electron-deuteron
scattering. The axial vector contribution comes mainly from the
strangeness form factors, as the radiative correction to the isoscalar
piece is quite small unlike the isovector case. Therefore, the
observation of ${\cal A}$ in elastic polarized electron deuteron
scattering is best suited to study the axial vector strangeness form
factor, if one has knowledge of the strangeness in the magnetic form
factor. In Figs.~5 and 6, we also show the effect of a
nonzero strangeness contribution in the magnetic and axial vector form
factors. We see that with the presently suggested value of $G_M^S$ and
$G_A^S$, the effect on ${\cal A}$ from $G_M^S$ and $G_A^S$ are
opposite in nature. The strangeness contribution to ${\cal A}$ is
dominated by $G_M^S$ as compared to $G_A^S$. This asymmetry was
discussed earlier by many authors~\cite{Po81,PO90, FrHM91,HwH80}. 
While Porrman~\cite{Po81} discusses ${\cal A}_z$ and ${\cal
  A}_\gamma$, he does not take into account the effect of a strangeness
content in the vector and axial vector currents, the work 
of Hwang and Henley~\cite{HwHM81} discusses only ${\cal A}_z$. On the
other hand, the work of Pollock~\cite{PO90} and Frederico et al.~\cite{FrHM91}
considers the asymmetry ${\cal A}_z$ by taking into account an
isoscalar axial vector contribution, while the effect of ${\cal
  A}_\gamma$ is neglected. Thus the present study is the first one,
which takes into account all contributions and presents results for
${\cal A}_z$ and ${\cal A}_\gamma$ for the electron energies of
current interest. 

The parity violation effects also enter in the tensor polarizations of
the recoil deuteron through $A_{ed}^{00+}(20+)$, $A_{ed}^{00+}(21+)$,
and $A_{ed}^{00+}(22+)$ which are given in terms of Sachs form factors
\begin{eqnarray}
S_0A_{ed}^{00+}(20+)&=&-\frac{8\sqrt{2}}{3}g_v^d\tilde{G_a}
\left[G_C+\frac{\eta}{3}G_Q\right]G_Q 
-\frac{\sqrt{2}}{3}\eta\left[1+2\left(1+\eta\right)
\tan^2\frac{\theta}{2}\right]g_v^d\tilde{G_a}G_M^2\nonumber\\
&&-\frac{2\sqrt{2}}{3}\eta\sec\frac{\theta}{2}\tan
\frac{\theta}{2}\sqrt{\left(1+\eta\right)
\left(1+\eta\sin^2\frac{\theta}{2}\right)}
\left[G_{E}+\tilde{G_v}G_A\right]G_M\,,\\
S_0A_{ed}^{00+}(21+)&=&\frac{8}{\sqrt{3}}\eta\sec
\frac{\theta}{2}\sqrt{\eta\left(1+\eta\sin^2\frac{\theta}{2}\right)}
g_v^d\tilde{G_a}G_MG_Q\nonumber\\
&&+\frac{4}{\sqrt{3}}\eta\tan\frac{\theta}{2}
\sqrt{\eta\left(1+\eta\right)}\left[G_{E}+\tilde{G_v}G_A\right]G_Q\,,\\
S_0A_{ed}^{00+}(22+)&=&-\frac{2}{\sqrt{3}}\eta g_v^d\tilde{G_a}G_M^2\,.
\end{eqnarray}
While $A_{ed}^{00+}(22+)$ is given solely in terms of the Standard
Model parameters (see Eq.~(114)) the other components involve the
isoscalar axial vector piece and the parity violating hadronic
interaction in the combination $G_{E}+\tilde{G_v}G_A$,
the same which occurs in the asymmetry ${\cal A}$. In order to extract
these one has to measure tensor polarization asymmetries. For
example, one may define the tensor polarization asymmetry ${\cal
  A}_{zz}^p$ by 
\begin{equation}
{\cal A}_{zz}^p=\frac{1}{h}\frac{p_{zz}(\uparrow)-p_{zz}
(\downarrow)}{p_{zz}(\uparrow)+p_{zz}(\downarrow)}
=\frac{A_{ed}^{00+}(20+)_{pv}}{A_{d}^{00+}(20+)_{pc+pv}}\,.
\end{equation}
In Fig.~7, we show the results for ${\cal A}_{zz}^p$ as a function of
$Q^2$ for various values of electron energy $E$ = 125, 200, 315, 361, 
 and 687 MeV. We also show the effect of including
 strangeness form factors in the magnetic as well as axial vector form
 factors. We note that ${\cal A}_{zz}^p$ remains constant for a large
 range of $Q^2$ values. This is not surprising since ${\cal A}_{zz}^p$ 
is a ratio of $A_{ed}^{00+}(20+)$ and $A_{d}^{00+}(20+)$, which have a
similar $Q^2$ dependence, though differing in magnitude by a large amount.

\section{Summary and Conclusions}
Parity violating electron scattering experiments are being done by
the SAMPLE, HAPPEX, G0 and A4 collaborations at MIT, JLAB and MAINZ
from proton, deuteron and $^{4}$He targets in
the energy region of a few hundered MeV. These experiments are expected
to answer questions about the strange form factors of the nucleon and
radiative corrections to the axial vector couplings which may also
lead to an understanding of anapole moments. 

In this paper we have examined parity violating observables in the
scattering of unpolarized and polarized electrons from unpolarized
deuterons. The parity violating observables receive contributions from
the $\gamma$-Z interference in the Standard Model and also from the
parity violating electromagnetic coupling which are induced 
by the parity violating components in the deuteron wave function. In
almost all parity 
violating observables, the contribution of the parity violating 
electromagnetic coupling is found to be small compared to the
$\gamma$-Z interference contribution. In the case of unpolarized electron
scattering, the nonvanishing components of the parity violating recoil
vector polarization of the deuteron have been studied at electron
energies $E$ = 125,
200, 315, 362, and 687 MeV, relevant for the ongoing experiments as a
function of $Q^2$. In addition, for a fixed value of $Q^2$, these 
polarizations have been studied as a function of the scattering angle. The
effects of a nonzero strangeness component in the magnetic form factor as
well as the axial vector form factor have been calculated. These
effects are found to be important in backword 
direction. In the case of the axial vector coupling, the radiative
corrections are known to be large in the isovector component and play
an important role in the analysis of quasi-elastic electron deuteron
scattering. For elastic electron deuteron scattering, where
isoscalar terms contribute, these radiative corrections are small, thus
making it a suitable method to study the strangeness form factor. 

In the case of polarized electron scattering from unpolarized deuterons,
the parity violating electron beam asymmetry and the tensor
polarization asymmetries of the recoiling deuteron have been
studied. For the electron beam asymmetry, the results are presented as
a function of $Q^2$ for various energies. Furthermore, the
asymmetries have been studied as a function of the scattering angle
at fixed $Q^2$. The 
effect of including a nonzero component in the magnetic form factor
$G_M^s(Q^2)$ as well as an axial form factor $G_A^s(Q^2)$ have been
investigated too. 

In all these cases, the contributions to the parity violating
observables are dominated by the $\gamma$-Z interference term of the 
Standard Model where the leptonic axial vector current interacts with
the hadronic isoscalar axial current. There are additional
contributions, though small, coming from the parity violating
electromagnetic coupling as well as from the axial vector isoscalar
hadronic current which interacts with the leptonic vector
current. This isoscalar hadronic current may be due to the strangeness
component of the nucleon in the Standard Model or due to the structure of
weak interactions beyond the Standard Model. It is found
that the parity violating observables are important in the backward
direction. The presented results for various parity violating
asymmetries might be helpful in analysing future experiments on
electron-deuteron scattering being done at MAINZ, MIT and JLAB. 
\section{Acknowledgment}
S.K.\ Singh would like to thank the Humboldt Foundation for financial
support during the course of this work and S.\ Ahmad would like to
thank CSIR, New Delhi, India for a research fellowship. H.\
Arenh\"ovel gratefully acknowledges the 
warm hospitality of the Physics Department of Aligarh Muslim
University and financial support by the Deutsche Forschungsgemeinschaft
(SFB 443).

\begin{figure}
\includegraphics[width=12cm]{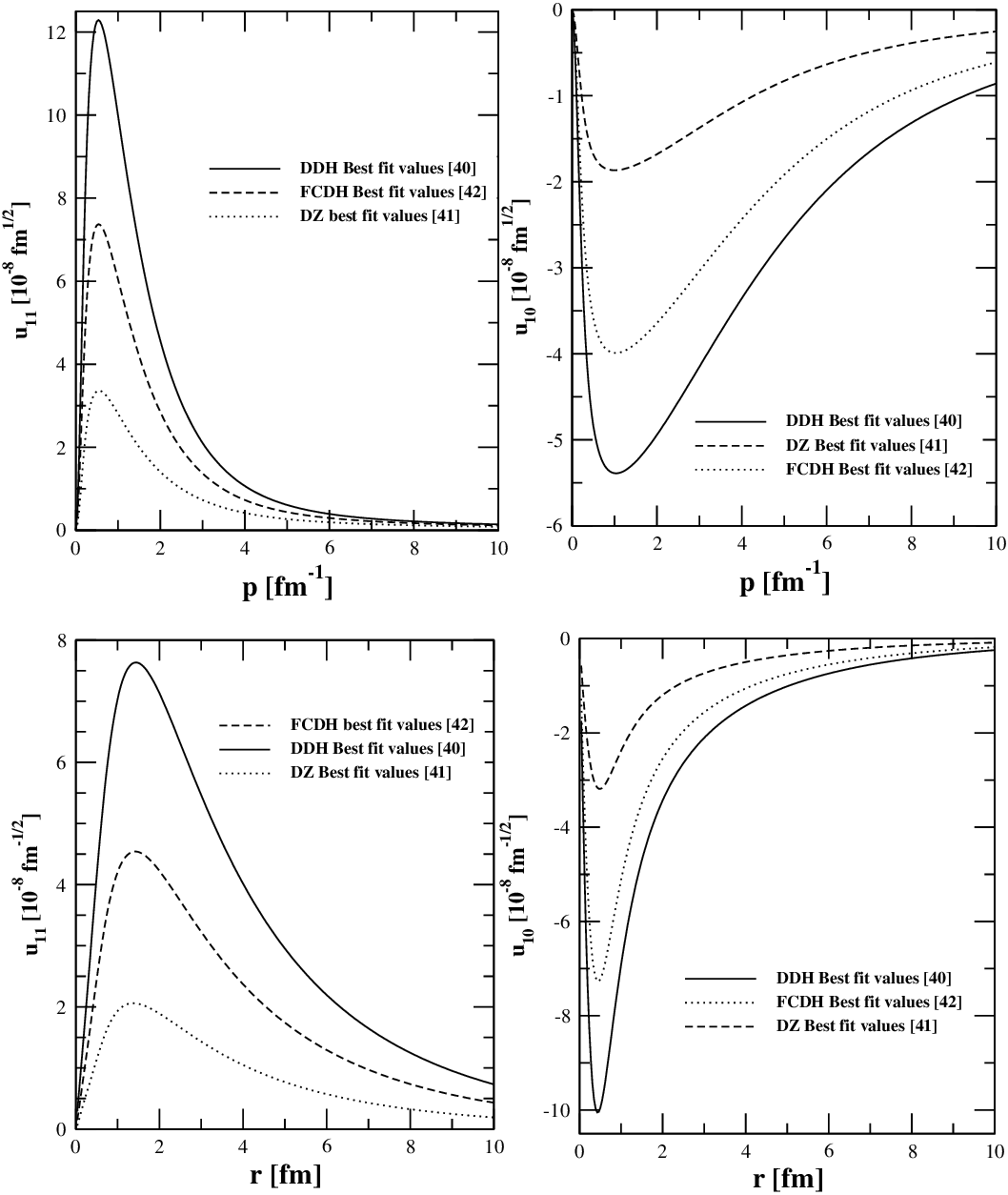}
\caption{Radial part of p-wave components of the deuteron wave
  function in momentum space and in configration space. Separately
  shown are the sensitivity to the parameters of the weak
  nucleon-nucleon potential.}
\end{figure}

\begin{figure}[t]
\includegraphics[width=14cm]{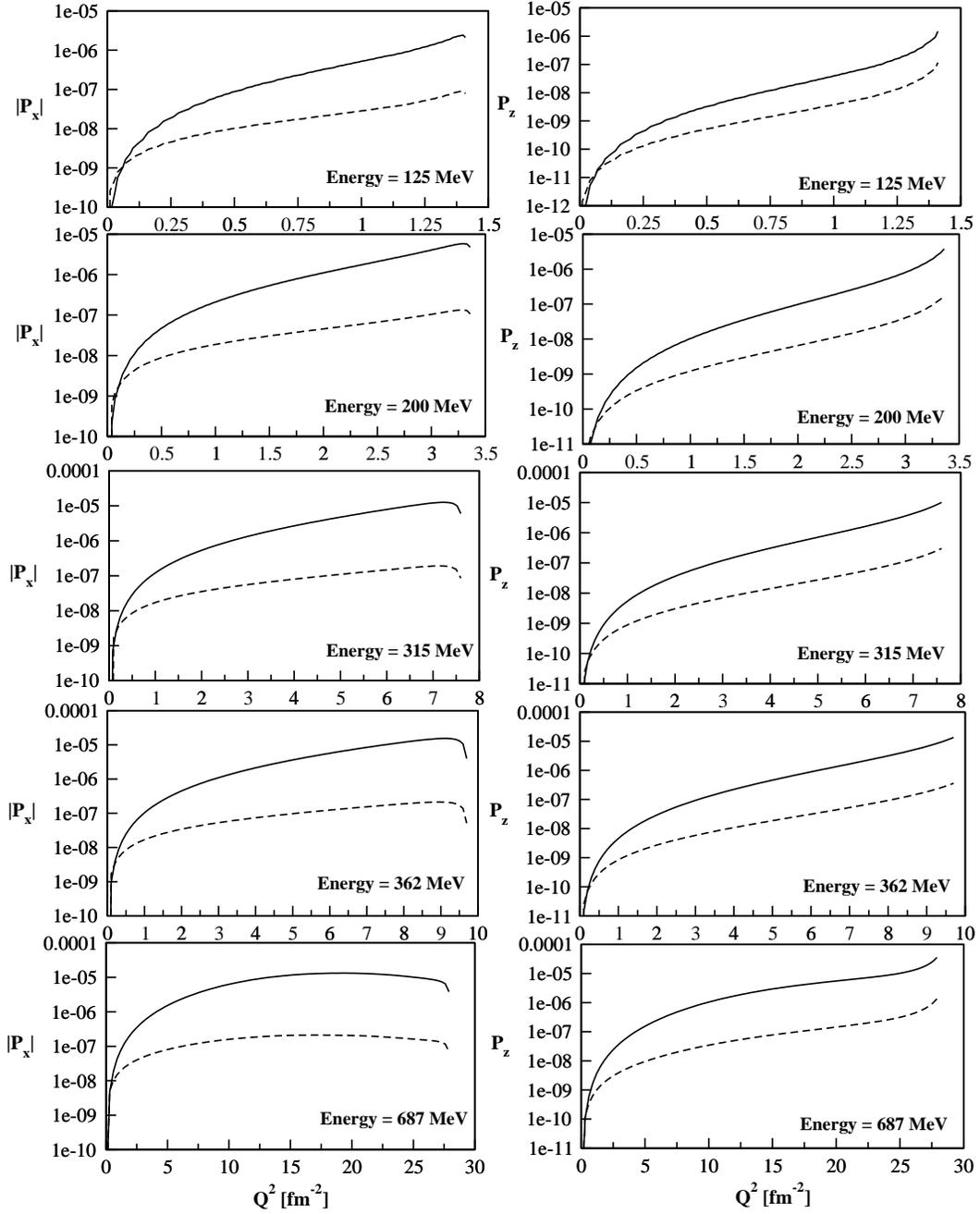}
\caption{Recoil vector polarizations $P_x$ (left panel) and $P_z$
  (right panel) as a function of $Q^2$ for various energies $E$ = 125,
  200, 315, 361, and 687 MeV. The solid curves represent the
  contribution of Z-exchange and the dashed curves the contribution of
  the $\gamma$-Z interference due to P state admixtures in the
  deuteron wave function.}  
\end{figure}
\begin{figure}[t]
\includegraphics[width=14cm]{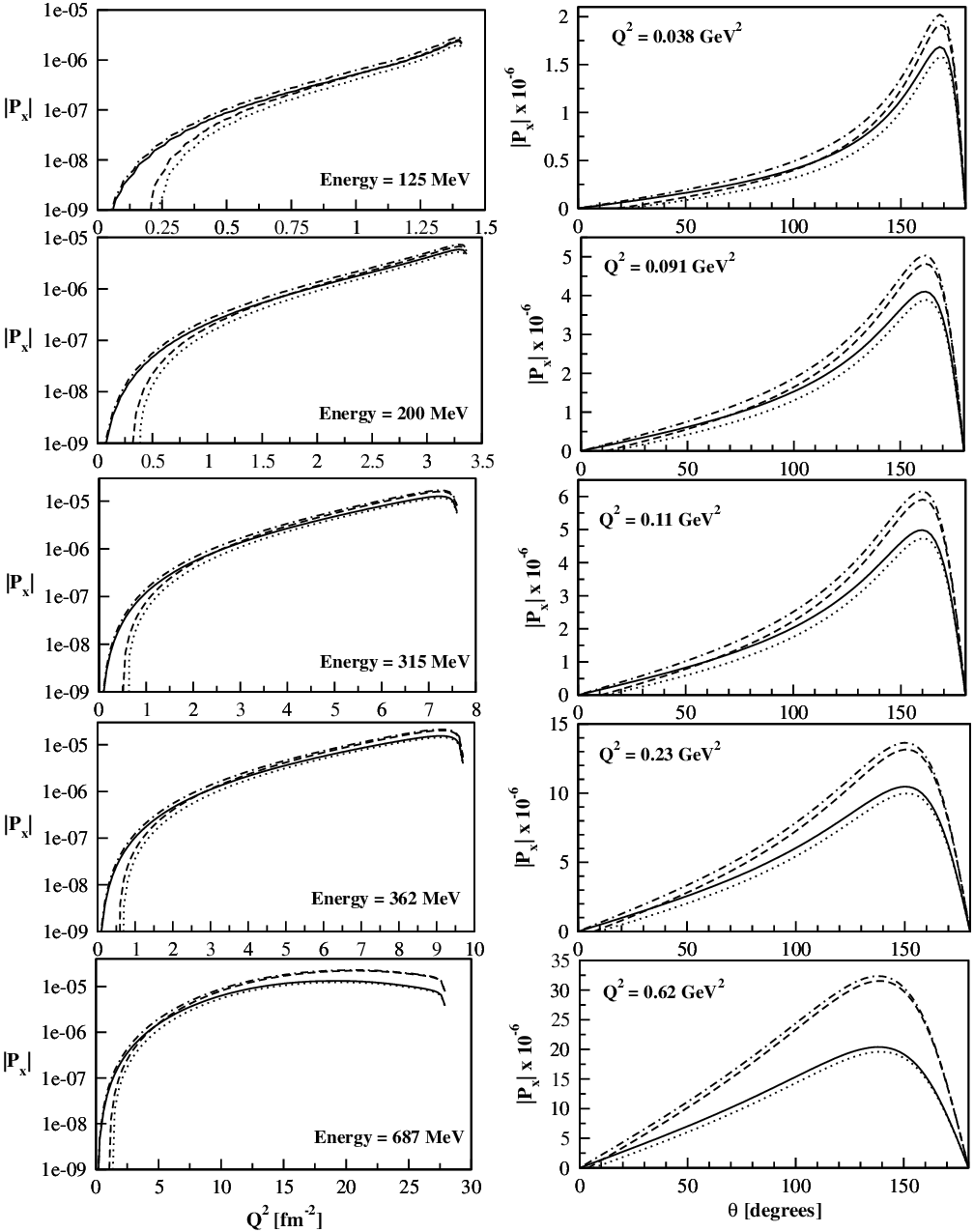}
\caption{Recoil vector polarization $P_x$ by Z-exchange as a function
  of $Q^2$ for various energies $E$ = 125, 200, 315, 361, and 687 MeV
  (left panel), and as a function of $\theta$ for fixed $Q^2$ = 0.038,
  0.091, 0.11, 0.23, and 0.62 $GeV^2$. The effect of strangeness form
  factors in magnetic as well as axial vector form factors are 
  shown for three cases: (i) $G_M^s\neq 0$ and $G_A^s=0$
  (dashed-dotted), (ii) $G_M^s=0$ and $G_A^s\neq 0$ (dotted), and
  (iii) $G_M^s\neq 0$ and $G_A^s\neq0$ (dashed). The solid curves
  represent the case when $G_M^s=0$ and $G_A^s=0$.} 
\end{figure}
\begin{figure}[t]
\includegraphics[width=14cm]{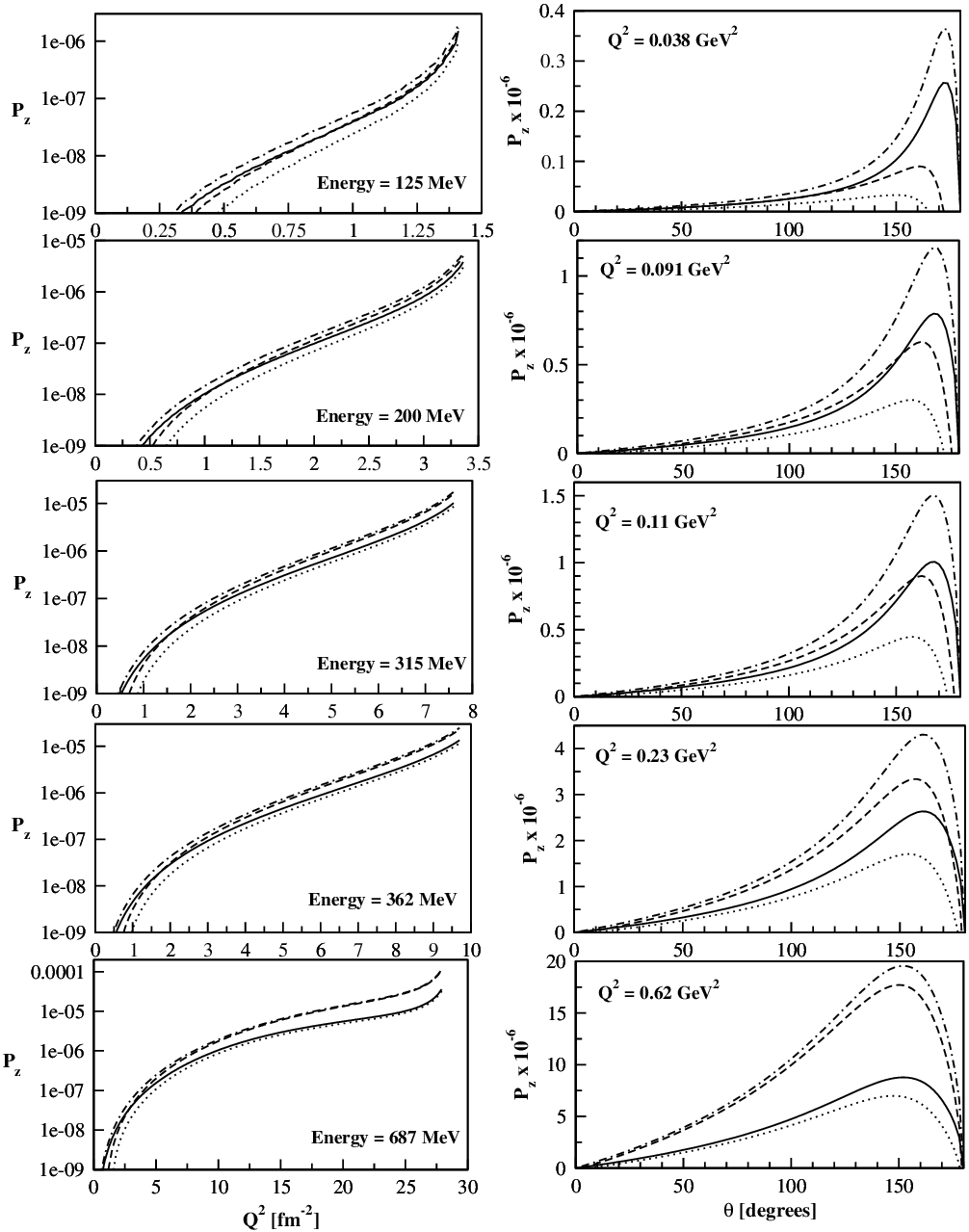}
\caption{Recoil vector polarizations $P_z$ by Z-exchange as a
  function of $Q^2$ for various energies $E$ = 125, 200, 315, 361, and
  687 MeV (left panel), and as a function of $\theta$ for fixed $Q^2$
  = 0.038, 0.091, 0.11, 0.23, and 0.62 GeV$^2$. The effect of
  strangeness form factors in magnetic as well as axial vector form
  factors are shown for three cases: (i) $G_M^s\neq 0$ and $G_A^s=0$
  (dashed-dotted), (ii) $G_M^s=0$ and $G_A^s\neq 0$ (dotted), and
  (iii) $G_M^s\neq 0$ and $G_A^s\neq0$ (dashed). The 
  solid curves represent the case when $G_M^s=0$ and $G_A^s=0$.} 
\end{figure}
\begin{figure}[t]
\includegraphics[width=14cm]{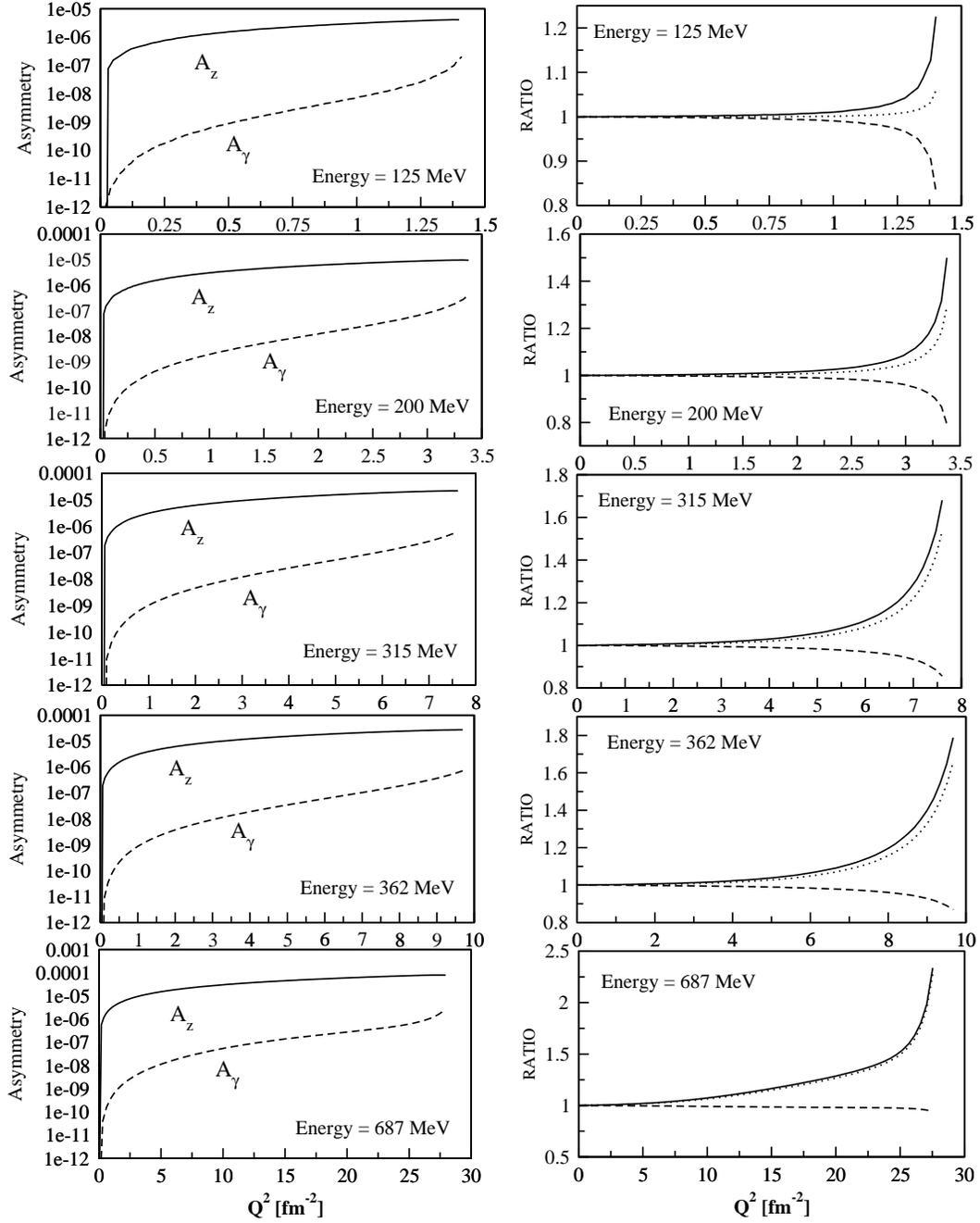}
\caption{Elastic deuteron asymmetry by Z [${\cal A}_z$ (solid)]
  and photon exchange [${\cal A}_\gamma$ (dashed)] as a function
  of $Q^2$ for various incident energies $E$ = 125, 200, 315, 361, and
  687 MeV (left panel). The effect of strangeness form factors in
  magnetic $G_M^s$ as well as axial vector $G_A^s$ form factor are
  shown as a ratio of strange contributions  (right panel): (i)
  $G_M^s\neq 0$ and $G_A^s=0$ (solid), (ii) $G_M^s=0$ and
  $G_A^s\neq 0$ (dashed), and (iii) $G_M^s\neq 0$ and $G_A^s\neq0$
  (dotted), with no strange contributions $G_M^s=0$ and
  $G_A^s=0$. } 
\end{figure}
\begin{figure}[t]
\includegraphics[width=14cm]{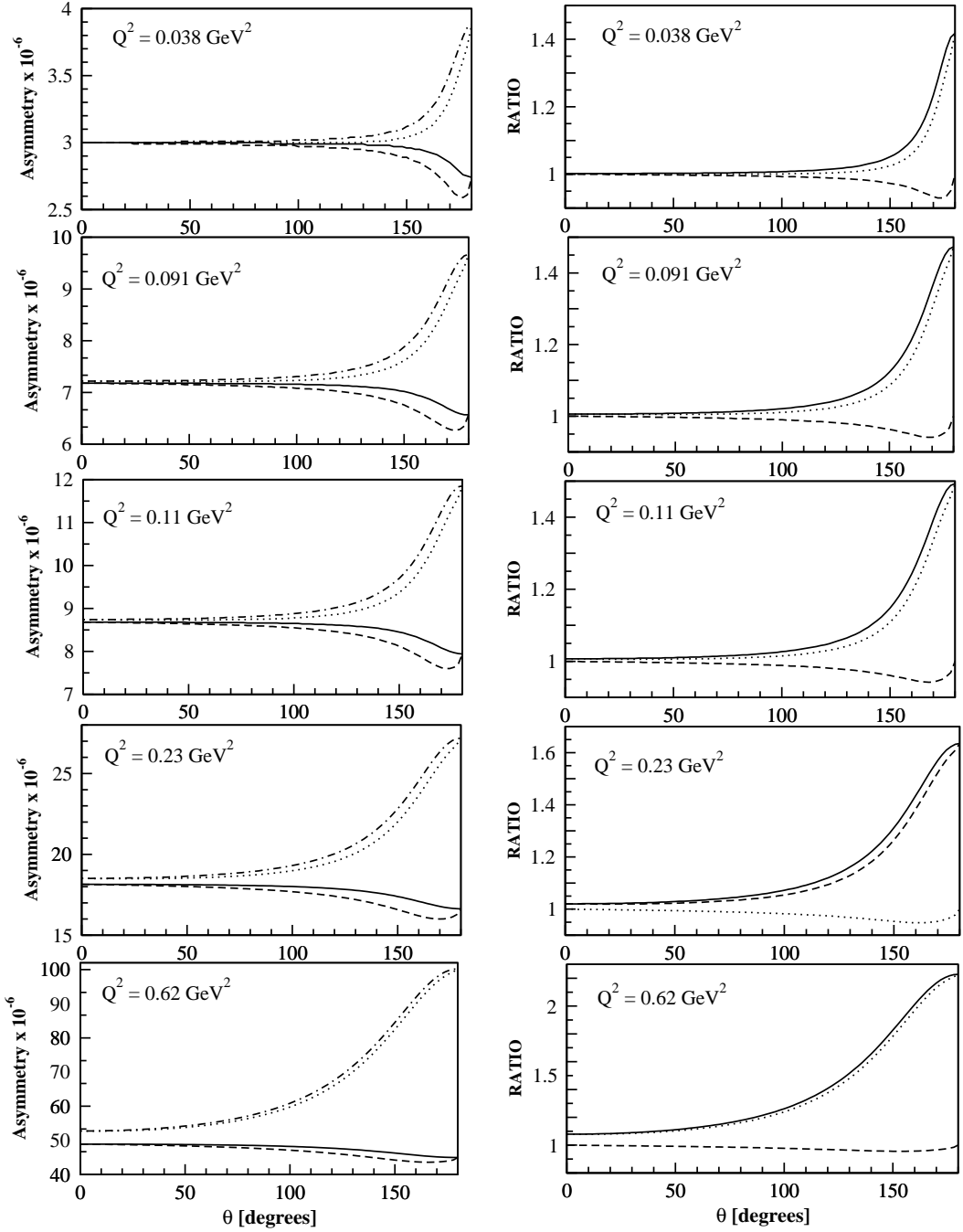}
\caption{Elastic deuteron asymmetry by Z-exchange [${\cal A}_z$
  (solid)] as a function of $\theta$ for fixed $Q^2$ = 0.038, 0.091, 
  0.11, 0.23, and 0.62 GeV$^2$ (left panel). The effect of strangeness
  form factors in magnetic as well as axial vector form factors have
  been shown for three cases: (i) $G_M^s\neq 0$ and $G_A^s=0$
  (dashed-dotted), (ii) $G_M^s=0$ and $G_A^s\neq 0$ (dashed), and
  (iii) $G_M^s\neq 0$ and $G_A^s\neq0$ (dotted). The 
  solid curves represent the case when $G_M^s=0$ and $G_A^s=0$. In the
  right panel of the figure the effect of strangeness form factors
  have been shown as a ratio of strange contributions (i) $G_M^s\neq
  0$ and $G_A^s=0$ (solid), (ii) $G_M^s=0$ and $G_A^s\neq 0$
  (dashed), and (iii) $G_M^s\neq 0$ and $G_A^s\neq0$ (dotted), with no
  strange contributions $G_M^s=0$ and $G_A^s=0$. } 
\end{figure}
\begin{figure}[t]
\includegraphics[width=14cm]{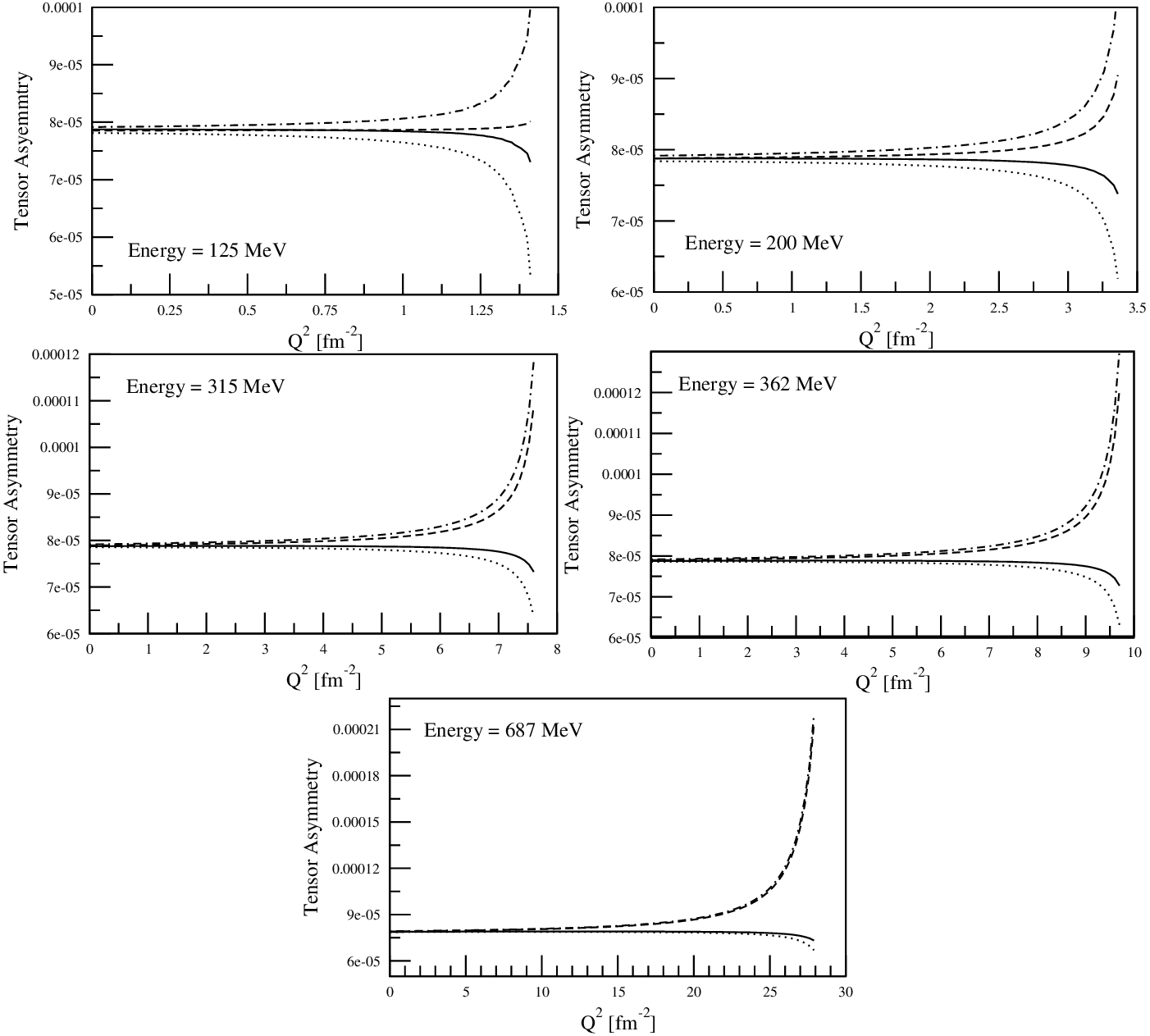}
\caption{Recoil tensor asymmetry ${\cal A}_{zz}^p$ from Z-exchange as a
  function of $Q^2$ for various incident energies $E$ = 125, 200, 315,
  361, and 687 MeV. The effect of strangeness form factors in magnetic
  as well as axial vector form factors are shown for three cases: (i)
  $G_M^s\neq 0$ and $G_A^s=0$ (dashed-dotted), (ii) $G_M^s=0$
  and $G_A^s\neq 0$ (dotted), and (iii) $G_M^s\neq 0$ and
  $G_A^s\neq0$ (dashed). The solid curves represent the case when
  $G_M^s=0$ and $G_A^s=0$.} 
\end{figure}

\end{document}